\pdfoutput=1
\documentclass[12pt]{article}
\let\mysub=\subset

\DeclareFontFamily{T1}{calligra}{}
\DeclareFontShape{T1}{calligra}{m}{n}{<->s*[1.44]callig15}{}
\DeclareMathAlphabet\mathcalligra   {T1}{calligra} {m} {n}
\DeclareMathAlphabet\mathzapf       {T1}{pzc} {mb} {it}
\DeclareMathAlphabet\mathchorus     {T1}{qzc} {m} {n}
\DeclareMathAlphabet\mathrsfso      {U}{rsfso}{m}{n}
\DeclareMathAlphabet\mathfrcal      {T1}{frcursive}{m}{it}
\DeclareFontFamily{T1}{frcursive}{}
\DeclareFontShape{T1}{frcursive}{m}{n}{<->s*[1.44]callig15}{}
\DeclareMathAlphabet\mathfrcal      {T1}{frcursive}{m}{it}

\usepackage{amsmath}
\usepackage{amssymb}
\usepackage{graphicx}
\usepackage{xcolor}
\numberwithin{equation}{section}
\usepackage{array}
\usepackage{mathtools}
\usepackage{dsfont}
\usepackage{mathrsfs}
\usepackage{tikz}\usetikzlibrary{matrix,fit}
\usepackage{varwidth}
\usepackage{enumerate}
\usepackage{appendix}

\usepackage[compat=1.1.0]{tikz-feynman}

\usepackage[margin=1in]{geometry}
\usepackage{nicematrix}

\usepackage[
    backend=bibtex,
    style=alphabetic,
maxbibnames=99,
giveninits=true,
minalphanames=1,
maxalphanames=3,url=false
  ]{biblatex}

  \bibliography{SymmSpaces}

\usepackage{mathabx}
\usepackage{empheq}

\usepackage{hyperref}
\usepackage{setspace}

\usepackage{slashed}
\usepackage{upgreek}
\usepackage{appendix}

\usepackage{tabstackengine}

\usepackage{wrapfig}
\usepackage[abs]{overpic}

\usepackage{float}

\fixTABwidth{T}

\newdimen\mytextwidth
\newcommand\rem[2][cyan!40!green]{\noindent\nobreak\hfil\penalty1000\hfilneg
\mytextwidth=\linewidth\advance\mytextwidth by 2mm
\begin{tikzpicture}[baseline=-\the\dimexpr\fontdimen22\textfont2\relax]\node[outer sep=0pt,draw=black,fill=#1,fill opacity=1,text opacity=1,rectangle,rounded corners]{\begin{varwidth}{\mytextwidth}\textcolor{white}{#2}\end{varwidth}};
\end{tikzpicture}\allowbreak
}

\newcommand\whiterem[2][white!]{\noindent\nobreak\hfil\penalty1000\hfilneg
\mytextwidth=\linewidth\advance\mytextwidth by 2mm
\begin{tikzpicture}[baseline=-\the\dimexpr\fontdimen22\textfont2\relax]\node[outer sep=0pt,draw=black,fill=#1,fill opacity=1,text opacity=1,rectangle,rounded corners,line width=1.5pt]{\begin{varwidth}{\mytextwidth}\textcolor{black}{#2}\end{varwidth}};
\end{tikzpicture}\allowbreak
}

\usepackage{accents}
\newcommand\thickbar[1]{\accentset{\rule{.7em}{.8pt}}{#1}}

\newcommand{\dd}{\partial}

\newcommand{\CP}{\mathsf{CP}}
\newcommand{\CC}{\mathsf{C}}

\newcommand{\bea}{\begin{equation}}
\newcommand{\eea}{\end{equation}}
\newcommand{\bear}{\begin{eqnarray}}
\newcommand{\eear}{\end{eqnarray}}
\newcommand{\bearr}{\begin{eqnarray*}}
\newcommand{\eearr}{\end{eqnarray*}}

\newcommand{\tSU}{\textbf{SU}}
\newcommand{\tGL}{\textbf{GL}}
\newcommand{\tU}{\textbf{U}}
\newcommand{\tO}{\textbf{O}}
\newcommand{\tSO}{\textbf{SO}}
\newcommand{\tSL}{\textbf{SL}}
\newcommand{\tSp}{\textbf{Sp}}
\newcommand{\tPSL}{\textbf{PSL}}

\renewbibmacro{in:}{}

\usepackage{mdframed}

\newbibmacro{string+doi}[1]{
  \iffieldundef{doi}{#1}{\href{http://dx.doi.org/\thefield{doi}}{#1}}}
\DeclareFieldFormat{title}{\usebibmacro{string+doi}{\mkbibemph{#1}}}
\DeclareFieldFormat[article]{title}{\usebibmacro{string+doi}{\mkbibquote{#1}}}

\newmdenv[
  topline=false,
  bottomline=false,
  rightline=false,
  linewidth=2pt,
  skipabove=\topsep,
  skipbelow=\topsep
]{siderules}

\newmdenv[
  topline=false,
  bottomline=false,
  linewidth=2pt,
  skipabove=\topsep,
  skipbelow=\topsep
]{siderulesright}

\makeatletter
\renewcommand{\@seccntformat}[1]{\csname the#1\endcsname.\quad}
\makeatother

\usepackage{setspace}
\usepackage{xpatch}

\makeatletter
\renewcommand{\@chap@pppage}{
  \clear@ppage
  \thispagestyle{plain}
  \if@twocolumn\onecolumn\@tempswatrue\else\@tempswafalse\fi
  \null\vfil
  \markboth{}{}
  {\centering
   \interlinepenalty \@M
   \normalfont
   \MakeUppercase \appendixpagename\par}
  \if@dotoc@pp
    \addappheadtotoc
  \fi
  \vfil\newpage
  \if@twoside
    \if@openright
      \null
      \thispagestyle{empty}
      \newpage
    \fi
  \fi
  \if@tempswa
    \twocolumn
  \fi
}
\makeatother

\definecolor{navycol}{RGB}{100,150,160}
   \definecolor{pinkcol}{RGB}{242,55,55}
   \definecolor{greencol}{RGB}{50,205,50}

   \definecolor{bluecol}{RGB}{30,144,255}

\usepackage{hyperref}

\usepackage{titlesec}

\titleformat*{\section}{\large\bfseries}
\titleformat*{\subsection}{\normalsize\bfseries}
\titleformat*{\subsubsection}{\normalsize\bfseries}
\titleformat*{\paragraph}{\large\bfseries}
\titleformat*{\subparagraph}{\large\bfseries}
\titlespacing{\author}{-5pt}{-5pt}{-5pt}[-5pt]

\makeatletter
\renewcommand\subsubsection{\@startsection{subsubsection}{3}{\z@}
                                     {-3.25ex\@plus -1ex \@minus -.2ex}
                                     {-1.5ex \@plus -.2ex}
                                     {\normalfont\normalsize\bfseries}}
\renewcommand\subsection{\@startsection{subsection}{3}{\z@}
                                     {-3.25ex\@plus -1ex \@minus -.2ex}
                                     {-1.5ex \@plus -.2ex}
                                     {\normalfont\normalsize\bfseries}}                                     
\makeatother

\DeclareFontFamily{U}{solomos}{}
\DeclareErrorFont{U}{solomos}{m}{n}{10}
\DeclareFontShape{U}{solomos}{m}{n}{
  <-> s*[1.1]  gsolomos8r
}{}

\newcommand{\vkappa}{\text{\usefont{U}{solomos}{m}{n}\symbol{'153}}}

   \usepackage{stmaryrd}

\usepackage{tikz}
\usetikzlibrary{arrows.meta}

\usepackage{indentfirst}

\usepackage{tocloft}
\let \savenumberline \numberline
\def \numberline#1{\savenumberline{#1.}}

\usepackage{etoolbox}
\patchcmd{\tableofcontents}{\@starttoc}{\vspace{-0.3cm}\@starttoc}{}{}

\begin{document}

\title{\vspace{-1.0cm} \textbf{Grassmannian Sigma Models}
\vspace{0.5cm}
}

\author{Dmitri Bykov\footnote{Emails:
 bykov@mi-ras.ru, dmitri.v.bykov@gmail.com} \qquad\qquad Viacheslav Krivorol\footnote{Emails:
 vkrivorol@itmp.msu.ru, v.a.krivorol@gmail.com}
\\  \vspace{0cm}  \\
{\small $\bullet$ \emph{Steklov
Mathematical Institute of Russian Academy of Sciences,}} \\{\small \emph{Gubkina str. 8, 119991 Moscow, Russia} }\\
{\small $\bullet$ \emph{Institute for Theoretical and Mathematical Physics,}} \\{\small \emph{Lomonosov Moscow State University, 119991 Moscow, Russia}}}

\date{}

{\let\newpage\relax\maketitle}

\maketitle

\vspace{0cm}
\textbf{Abstract.} We show that sigma models with  orthogonal and symplectic Grassmannian target spaces admit chiral Gross-Neveu model formulations, thus extending earlier results on unitary Grassmannians.  
As a first application, we calculate the one-loop $\beta$-functions in this formalism, showing that they are proportional to the dual Coxeter numbers of the respective symmetry algebras.

\tableofcontents

\vspace{0.5cm}

\section{Introduction}

For a long time the study of integrable 2D sigma models concentrated on the case of symmetric target spaces (cf. the books~\cite{AbdallaBook, GuestBook, HarmonicMapsBook} and the review~\cite{PerelomovChiral} for a geometric background). It was later observed that there are broad classes of non-symmetric target spaces (such as flag manifolds), for which integrability is apparently present as well\footnote{In the present paper we focus on homogeneous models. Another generalization is related to integrable (non-homogeneous) deformations of these models -- a topic actively studied in the past years, cf.~\cite{FOZ, Klimcik0, Lukyanov, Klimcik, DelducVicedo, Sfetsos, BazhanovKotousov, FateevLitvinov, HoareTseytlin, Alfimov, KotousovTeschner, LiniadoVicedo} and references therein.}. It was argued in~\cite{BykovComplex} that it is perhaps the existence of complex structure on the target space that is the key property ensuring integrability in this class of models. Subsequently~\cite{CY} emphasized that the crucial ingredient is the complex symplectic structure on the phase space of the model (which could be thought of as the cotangent bundle to the target space). It is quite likely, though, that at the quantum level not an arbitrary complex symplectic phase phase would be admissible. It was then proposed in~\cite{BykovGN} that one should restrict to models admitting gauged linear (GLSM) descriptions. In physics language these are nothing but bosonic incarnations of generalized chiral Gross-Neveu (GN) models~\cite{GrossNeveu, WittenThirring} (see also~\cite{Anselm} and the discussion in~\cite{ShifmanAnselm}), which could be called the `$\phi^4$-theory in 2D'. As for the math interpretation, these are models whose phase spaces are quiver varieties \cite{Nakajima}. We believe that it is this relation to Gross-Neveu models, which were solved long ago~\cite{ALGrossNeveu, DestriDeVega}, that provides the simplest explanation of integrability of these models.

The class of models that have been recast in the form of GN-models so far included theories with $\tSU(n)$-symmetry: $\CP^{n-1}$, Grassmannians and flag manifolds~\cite{BykovHermitian, BykovSUSY, BykovGN}. A natural expectation is that a similar program could be carried out for theories with other symmetry groups. The present paper takes a step in this direction, concentrating on families of orthogonal (symmetry group $\tO(n)$) and symplectic (symmetry group\footnote{Here by $\tO$ and $\tSp$ we mean the compact forms 
\bea
\tO(n)=\tO(n, \CC)\bigcap \tU(n)\quad  \textrm{and}\quad  \tSp(2n)=\tSp(2n, \CC)\bigcap \tU(2n)\,.
\eea
Complex groups are generally referred to as $\tGL(n, \CC)$, $\tO(n, \CC)$ and $\tSp(2n, \CC)$.} $\tSp(2n)$) Grassmannians. Including the $\tSU(n)$-case for completeness, these families are\footnote{Note that $\mathrm{\mathsf{OGr}}(m, 2m)$ has two connected components.   
We will discuss it in detail in Appendix~\ref{OGrHomSpacesec} below.}:

\begin{empheq}[box = \fcolorbox{black}{white}]{align}
 \label{UGr}
&\quad \textrm{Unitary:}\;\;\quad\quad\quad \mathsf{Gr}(m, n):=\frac{\tU(n)}{\tU(m)\times \tU(n-m)}\,,\\ \label{OGr}
&\quad\textrm{Orthogonal:}\quad\quad 
\mathsf{OGr}(m, n):=\frac{\tO(n)}{\tU(m)\times \tO(n-2m)}\,,
\\ \label{SGr}
&\quad\textrm{Symplectic: }\quad\quad \mathsf{SGr}(m, 2n):=\frac{\tSp(2n)}{\tU(m)\times \tSp(2(n-m))}\,. \quad
\end{empheq}

\vspace{0.3cm}
Though symmetric space models are only part of the full landscape of integrable sigma models, they do stand out. The reason is that it is only for these spaces that our construction produces K\"ahler metrics for the sigma models. In fact, all of the Grassmannians above admit K\"ahler metrics\footnote{K\"ahler homogeneous spaces were classified in~\cite{Borel}.}, but in general these are different from the so-called `normal', or reductive, metric that our construction produces. 
It is likely that for other choices of metrics integrability of the models is lost. When complex structure is present, the relevant symmetric spaces are the  
Hermitian symmetric spaces.  In Table~\ref{SymmetricSpaces} we explicitly write out those of the Grassmannians above that are Hermitian symmetric\footnote{In fact, these exhaust all Hermitian symmetric spaces of the classical groups.} (in the last entry we keep one connected component of $\mathsf{OGr}(n, 2n)$).
\begin{table}[h]
\centering
\begin{tabular}{|cc|}
\hline
\multicolumn{1}{|c|}{Hermitian symmetric space} & Name and interpretation \\ \hline
                        &   \\
$\displaystyle\frac{\tU(n)}{\tU(m)\times \tU(n-m)}$                    & \raisebox{1 em}{Unitary Grassmannians} \\
                        & \raisebox{1 em}{$m=1, \ldots, n-1$}  \\ \hline
                        &   \\      
$\displaystyle\frac{\tSp(2n)}{\tU(n)}$                       & \raisebox{1 em}{Lagrangian Grassmannian} \\
                        & \raisebox{1 em}{Compact analog of the Siegel upper half space}  \\ \hline
                        &   \\        
$\displaystyle\frac{\tSO(n)}{\tSO(n-2)\times \tSO(2)}$                       & \raisebox{1 em}{Grassmannian of real 2-planes} \\
                        & \raisebox{1 em}{Quadric in $\CP^{n-1}$}  \\ \hline
                        &   \\
$\displaystyle\frac{\tSO(2n)}{\tU(n)}$                       & \raisebox{1 em}{Space of orthogonal complex structures} \\
                        & \raisebox{1 em}{Variety of projective pure spinors (cf.~\cite{BerkovitsNekrasov})}  \\ \hline
\end{tabular}
\caption{\label{SymmetricSpaces}Hermitian symmetric spaces of the classical groups.}
\label{hermsymmspace}
\end{table}
We should point that GLSM's for all Hermitian symmetric spaces were constructed in~\cite{Higashijima}. As a partial generalization, GLSM's for general $\tU(n)$ flag manifolds were described in~\cite{DonagiSharpe}.

\vspace{0.3cm}
The paper is organized as follows. For each of the three families in~(\ref{UGr})-(\ref{SGr}) we construct the relevant GN formulations. 
The three cases are considered in sections~\ref{unitarygrass}, \ref{orthogonalgrass} and \ref{symplecticgrass}, respectively. In section~\ref{gaugeAzero} we explain how the gauge field may be eliminated in all of these models, leaving behind ungauged generalized GN models. In section~\ref{betafuncsec} we present the calculation of the one-loop beta functions of these models, checking that they match with the known results in the symmetric space case.

 In Appendix~\ref{homspaceapp} we recall the relation between two definitions of orthogonal and symplectic Grassmannians: the complex-geometric definition and the one in terms of homogeneous spaces.  
 In Appendix~\ref{invmetrapp} we discuss the construction of most general invariant metrics on the Grassmannians, showing that in the non-symmetric cases there are two-parametric families of metrics. In Appendix~\ref{alphagaugeapp} we analyze the dependence of the  sigma model Lagrangians on the choice of generalized `$\alpha$-gauge', akin to the one customary in Yang-Mills theory. 
 Finally, in Appendix~\ref{CP1betaapp} we check our general results for the one-loop $\beta$-functions in the simple special cases when the target space is the sphere $\CP^1$.

\section{Unitary Grassmannians}\label{unitarygrass}

We start by recalling the formulation of Grassmannian $\mathsf{Gr}(m, n)$ sigma models as gauged chiral Gross-Neveu models~\cite{BykovHermitian}: to this end we will first describe the relevant Gross-Neveu model and then prove its equivalence to the sigma model. The target space $\mathsf{Gr}(m, n)$ should be thought of  as the `configuration space' of the sigma model, and the crucial step will be to construct the Hamiltonian formulation with `phase space' $\mathrm{T}^\ast\mathsf{Gr}(m, n)$. It is convenient to represent this Hamiltonian system in terms of the enlarged `phase space' $\mathrm{T}^\ast\mathrm{Hom}(\CC^m, \CC^n)\simeq\CC^{2mn}$ subjected to some constraints arising from the procedure of symplectic reduction. 
The constructed first order theory with symplectic quotient `phase space' is, in essence, the Gross-Neveu formulation. This representation reflects the fact that the Grassmannian is a quiver variety \cite{Nakajima}. As we will show in the forthcoming sections, this strategy also works in the case of orthogonal and symplectic Grassmannians. We shall also explain that this construction implies a special form of the Hamiltonian, or more concretely $\mathsf{H}\sim\mathrm{Tr}(\mu\thickbar{\mu})$, where $\mu$ is the moment map for the residual global symmetry (after symplectic reduction).

To demonstrate how this scheme works in practice, introduce a set of matrix-valued fields
\bea
\label{UVHOM}
U\in \mathrm{Hom}(\CC^m, \CC^n),\quad\quad V\in  \mathrm{Hom}(\CC^n, \CC^m)\,,
\eea
as well as a gauge field $\mathcal{A}$ for the natural action 
$U\rightarrow Ug$, $V\rightarrow g^{-1}V$
of $\tGL(m, \CC)$. To relate to the GN model, we package these fields in a Dirac spinor $\Psi:=\begin{pmatrix}
   U \\ \thickbar{V} 
\end{pmatrix}$, where bar stands for Hermitian conjugate. The free part of the system is then described simply by the Dirac Lagrangian
\bea\label{UVGrassLagr}
\mathcal{L}_0=\mathrm{Tr}\left(\thickbar{\Psi}\slashed{D}\Psi\right)=\mathrm{Tr}\left(V \thickbar{D}U\right)+\mathrm{Tr}\left(\thickbar{U} D\thickbar{V}\right)\,,\quad\quad \thickbar{D}U=\thickbar{\dd}-U\thickbar{\mathcal{A}}\,.
\eea
We work in complex coordinates $(z,\thickbar{z}\,)=(x+iy,x-iy)$ and the complex derivatives $\partial$ and $\thickbar{\partial}$ are defined in the standard way\footnote{In the two-dimensional $(x,y)$-plane  the complex  derivatives (also called Wirtinger derivatives) are $\partial=\frac{1}{2}(\partial_x-i\partial_y)$ and $\thickbar{\partial}=\frac{1}{2}(\partial_x+i\partial_y)$.}. 
Thus,~(\ref{UVGrassLagr}) is a bosonic incarnation of a free fermion, also known in CFT language as the `symplectic boson' (cf.~\cite{Goddard}). Just as in the fermionic case, this system has a holomorphic current
\bea\label{KMcurr}
{1\over 2\pi}\,J_{\mathsf{sl}}:=UV\,,\quad\quad \thickbar{\dd} J_{\mathsf{sl}}=0\,.
\eea
Notice that variation of~(\ref{UVGrassLagr}) w.r.t. the gauge field~$\thickbar{\mathcal{A}}$ produces the constraint $VU=0$, so that indeed $J_{\mathsf{sl}}$ takes values in $\mathsf{sl}(n)$, as the notation suggests. Upon quantization, the modes of $J_{\mathsf{sl}}$ form the $\mathsf{sl}(n)$ Kac-Moody algebra.

There is a beautiful geometric way of interpreting the system~(\ref{UVGrassLagr}), as well as its interacting generalizations. To explain it, let us concentrate on the holomorphic piece of the Lagrangian, and carry out a mechanical reduction, so that $\thickbar{\dd}={d\over dt}$. If one omits the gauge field, what remains is just the pull-back to the worldline of the canonical one-form on $\mathrm{T}^\ast \mathrm{Hom}(\CC^m, \CC^n)$. This one-form, which we call $\uptheta$, is of `$pdq$'-type, its derivative being the symplectic form: $d\uptheta=\mathrm{Tr}\left(dU \wedge dV\right)$. The role of the gauge field is to perform symplectic reduction w.r.t. $\tGL(m, \CC)$. The relevant moment map constraint is obtained by varying the Lagrangian w.r.t. $\thickbar{\mathcal{A}}$: $VU=0$. Besides, there is a residual $\tPSL(n, \CC)$ global symmetry, whose respective  moment map is $\mu_{\mathsf{sl}}=UV$. Notice that
\bea\label{munilpeq}
\mu_{\mathsf{sl}}^2=0\,,
\eea
as a consequence of the constraint. Clearly, $\mu_{\mathsf{sl}}:={1\over 2\pi}\,J_{\mathsf{sl}}$, so that in quantum theory the moment map is essentially the Kac-Moody current of the free system. We shall keep the relative factor of $2\pi$ to distinguish between the two, but otherwise we will use them interchangeably.

In all of the models in the present paper interactions are introduced in a uniform manner~\cite{BykovGN}: one replaces the free action $\mathcal{S}_0=\int_{\Sigma}\,d^2z\,\mathcal{L}_0$  with\footnote{Generally, $\Sigma$ is a Riemann surface, but since in this paper we do not discuss global aspects, one can assume  $\Sigma\simeq\mathsf{C}$.}
\bea\label{actiongenstruct}
\mathcal{S}=\mathcal{S}_0+{\vkappa\over 2\pi}\,\int_{\Sigma}\,d^2z\,\mathrm{Tr}\left(J\thickbar{J}\,\right)\,,
\eea
where $J$ is the Kac-Moody current of the relevant symmetry group and $\vkappa$ is the coupling constant. For example, taking the basic Lagrangian~(\ref{UVGrassLagr}) and the current~(\ref{KMcurr}), we get the action\footnote{In normalizing the kinetic and interaction terms of the Lagrangian (\ref{unitaryaction}) we follow the conventions of \cite{BykovBetaFunction}. In the orthogonal and symplectic cases below one will have two vertices instead of one, each with a factor of $2\pi\vkappa$ in front of it, which means that in the  action~(\ref{actiongenstruct}) $\vkappa/2\pi$ should be replaced with $\vkappa/4\pi$ in those cases, see section \ref{betafuncsec}.}
\bea\label{unitaryaction}
\mathcal{S}=\int_{\Sigma}\,d^2z\,\left[\,2\Big(\mathrm{Tr}\left(V\thickbar{D}U\right)+\mathrm{Tr}\left(\thickbar{U}D\thickbar{V}\right)\Big)+{2\pi\vkappa }\,\mathrm{Tr}\left(\thickbar{U}UV\thickbar{V}\right) \right]\,.
\eea

For $\vkappa \neq 0$ we may use the e.o.m. to express $V$ and $\thickbar{V}$ in terms of $U, \thickbar{U}$ and their derivatives. Substituting back in the action, we get the sigma model Lagrangian
\bea\label{UGrassLagr}
\mathcal{L}={4\over 2\pi\vkappa}\,\mathrm{Tr}\left[\frac{1}{\thickbar{U}U}\,D\thickbar{U}\thickbar{D}U\right]\,.
\eea
This is gauge invariant w.r.t. $\tGL(m, \CC)$ gauge transformations $U\to U g$, $g\in \tGL(m, \CC)$.  We may use this gauge symmetry to set a partial gauge $\thickbar{U}U=\mathds{1}_m$. Expressing the gauge fields $\mathcal{A}$ and $\thickbar{\mathcal{A}}$ via their e.o.m. and substituting back in the Lagrangian, we find the familiar expression for the Fubini-Study metric (more exactly, its Grassmannian generalization for $m>1$). One also sees that $\vkappa^{-1}$ plays the role of the squared radius of the target space. As a result, the expansion in powers of $\vkappa$ matches the conventional $\alpha'$ sigma model perturbation theory, cf.~\cite{Polchinski} (expansion in the curvature of the target space).

In what follows we will treat~(\ref{UGrassLagr}) as the reference Lagrangian for comparing our normalizations of the metrics and coupling constants.

\section{Orthogonal Grassmannians $\mathsf{OGr}(m, n)$} \label{orthogonalgrass}

We recall the definition of orthogonal Grassmannians. Consider the vector space $\CC^n$ with a non-degenerate symmetric tensor $\mathds{h}_n$ on it. $\mathsf{OGr}(m, n)$ may be defined as the set of $m$-planes, isotropic w.r.t. $\mathds{h}_n$. Henceforth we package the $m$ vectors in a single $n\times m$ matrix $U$, so that the isotropy constraint takes the form $U^t \mathds{h}_nU=0$. Thus, $\mathsf{OGr}(m, n)$ is naturally embedded in the standard Grassmannian $\mathsf{Gr}(m, n)$. We will now develop an alternative formulation, based on complex quotients, which is necessary for constructing the GN-type model in this case.

\subsection{Grassmannian of real 2-planes.} 
Before passing to general  orthogonal Grassmannians let us start with the simplest one, $\mathsf{OGr}(1,n)$. 
As the matter field consider the doublet $W\in \mathrm{Hom}(\CC^2, \CC^n)$, defined as follows:
\begin{equation}
\label{W}
W=\begin{pmatrix}
U&V^t
\end{pmatrix}\,.
\end{equation}
To relate to the unitary case discussed above, one can imagine that we have packaged the $U$- and $V$-fields of that model in a single field $W$. Clearly, there is a natural right action of $\tSL(2, \CC)$ on $W$. However, as the gauge group of the model 
one should rather take the subgroup of upper-triangular matrices,  isomorphic to  $\CC^\ast \ltimes \CC$. The holomorphic piece of the free Lagrangian is\footnote{In certain affine cases  such as the conifold (when one has no $\CC^\ast$ gauge symmetry)  a similar free system was discussed in~\cite{GrassiPolicastro}.} 
\bea\label{real2planeLagr}
\mathcal{L}=\mathrm{Tr}\left(\thickbar{D}W\omega_2 W^t \mathds{h}_n\right),\quad \thickbar{D}W:=\thickbar{\dd}W-W\,\begin{pmatrix}
\thickbar{\mathcal{A}}&\thickbar{\mathcal{A}_+}\\
0& -\thickbar{\mathcal{A}}
\end{pmatrix}\,,\quad\omega_{2}:=\begin{pmatrix}
0&1\\
-1& 0
\end{pmatrix}\,,
\eea
whereas the constraints imposed by the gauge fields are
\bea\label{momapsl3}
V\mathds{h}_nU=0,\quad\quad U^t \mathds{h}_n U=0\,.
\eea
Again the residual symmetry group is $\tSO(n, \CC)$, with the moment map
\bea
\mu_{\mathsf{o}}=W\omega_2 W^t \mathds{h}_n=\left(U\otimes V-V^t\otimes U^t\right) \mathds{h}_n\,.
\eea
Notice that 
\bea\label{munilpeq2}
\mu_{\mathsf{o}}^2=-\left(V \mathds{h}_n V^t\right)\,U\otimes U^t \mathds{h}_n\,,
\quad \quad \textrm{and} \quad\quad \mu_{\mathsf{o}}^3=0\,.
\eea
The fact that $\mu_{\mathsf{o}}$ is nilpotent is parallel to the observation~(\ref{munilpeq}) in the unitary case.

\subsection{Orthogonal Grassmannians $\mathsf{OGr}(m,n)$.} 
To generalize to arbitrary $m$, one starts with pairs of canonical variables $U$ and $V$ both of which are $m\times n$ matrices, i.e.
\bea
U\in \mathrm{Hom}(\CC^m, \CC^{n}),\quad V\in  \mathrm{Hom}(\CC^{n}, \CC^m)\,.
\eea
As we already know from section~\ref{unitarygrass}, the choice of the gauge group $\tGL(m, \CC)$ would lead us to the unitary Grassmannians. Here instead $\tGL(m, \CC)$ will only be part of the gauge group. Extending the results of the previous section, we replace the second constraint in~(\ref{momapsl3}) by an $m\times m$ matrix of constraints:
\bea
\label{orthocons} H_{\mathrm{symm}}:=U^t \mathds{h}_n U=0\,,
\eea
as well as a symplectic reduction w.r.t. these constraints. Here  $H_{\mathrm{symm}}$ is  an  $m\times m$ symmetric matrix, $H_{\mathrm{symm}}^t=H_{\mathrm{symm}}$, 
transforming homogeneously under the action of $\tGL(m, \CC)$, i.e. $H_{\mathrm{symm}}\to g^t H_{\mathrm{symm}} g$. Therefore the  group of symplectic reduction is
\bea
\mathcal{G}=\tGL(m, \CC) \ltimes \mathrm{Mat}_{m}^{\mathrm{symm}}(\CC)\,,
\eea
where the second factor is the space of symmetric matrices, viewed as an additive abelian group\footnote{The matrix realization of this gauge group is
\bea
\begin{pmatrix}
g&0\\
0& (g^{-1})^t
\end{pmatrix}\begin{pmatrix}
1&\alpha\\
0& 1
\end{pmatrix}=\begin{pmatrix}
g&g\cdot\alpha\\
0& (g^{-1})^t
\end{pmatrix},\quad g\in \tGL(m, \CC),\quad \alpha^t=\alpha\,.
\eea
}. 
The dimension of the complex `space of fields' obtained after symplectic reduction is
\bea
\mathrm{dim}_{\CC}\Phi=2mn-2\,\mathrm{dim}_{\CC}\,\mathcal{G}=2\left(mn-{3m^2+m\over 2}\right)=2\, \mathrm{dim}_{\CC}\mathsf{OGr}(m, n)\,.
\eea
This is twice the dimension of the orthogonal Grassmannian, confirming the interpretation of the phase space as the cotangent bundle.

The Grassmannian of real 2-planes discussed above corresponds to $m=1$. The concrete Lagrangian generalizing~(\ref{real2planeLagr}) can be formulated using the same doublet (\ref{W}) $W\in \mathrm{Hom}(\CC^{2m}, \CC^n)$, where now $U$ and $V$ are $n\times m$ and $m\times n$ matrices respectively:
\bear\label{orthoGrassLagr}
&&\mathcal{L}=\mathrm{Tr}\left( \thickbar{D}W\omega_{2m}W^t\mathds{h}_n\right),\quad\quad\\ \nonumber &&\textrm{where}\quad\quad  \thickbar{D}W:=\thickbar{\dd}W-W\,\begin{pmatrix}
\thickbar{\mathcal{A}}&\thickbar{\mathcal{A}_+}\\
0& -\thickbar{\mathcal{A}}^t
\end{pmatrix},
\eear
with $\thickbar{\mathcal{A}}$ is a $\tGL(m, \CC)$ gauge field, $\thickbar{\mathcal{A}_+}$ a symmetric $m\times m$-matrix representing an abelian $\mathrm{Mat}_{m}^{\mathrm{symm}}(\CC)$  gauge field, and $\omega_{2m}=\begin{pmatrix}
0&\mathds{1}_m\\
-\mathds{1}_m& 0
\end{pmatrix}$ the reference symplectic form on $\CC^{2m}$. 
Writing this Lagrangian out in components,
\bea\label{orthofree}
\mathcal{L}\simeq 2\,\mathrm{Tr}\left(V\mathds{h}_n\thickbar{D}U\right)+
\mathrm{Tr}\left(\thickbar{\mathcal{A}_+} U^t\mathds{h}_nU\right)\,,\qquad \thickbar{D}U=\thickbar{\dd}U-U\thickbar{\mathcal{A}}\,,
\eea
where here and below the symbol $\simeq$ means equality up to integration by parts.
Variation w.r.t. $\thickbar{\mathcal{A}}$ leads to the constraint
\bea\label{VgU}
V\mathds{h}_nU=0\,,
\eea
whereas variation w.r.t. $\thickbar{\mathcal{A}_+}$ produces~(\ref{orthocons}). This is a straightforward higher-$m$ generalization of the constraints~(\ref{momapsl3}).

Clearly, the constraints preserve $\tSO(n, \CC)$ symmetry. The corresponding moment map is 
\begin{empheq}[box = \fcolorbox{black}{white}]{align}
\label{SOmommap2}
\quad \mu_{\mathsf{o}}=\left(U V-V^t U^t\right) \mathds{h}_n\,.\quad 
\end{empheq}

One reason we have kept the dependence on $\mathds{h}_n$, instead of immediately setting $\mathds{h}_n=\mathds{1}_n$, is to be able to eventually compare this expression with an analogous one in the symplectic case, see~(\ref{MomentMapSp}) below. For various calculations different choices of $\mathds{h}_n$ might be appropriate, but for technical simplicity we will always assume that $\mathds{h}_n$ is real and satisfies  $\mathds{h}_n^2=\mathds{1}_n$ (equivalently, $\mathds{h}_n\thickbar{\mathds{h}}_n=\mathds{1}_n$).

\subsection{Sigma model metrics on orthogonal Grassmannians.} General orthogonal Grassmannians are not symmetric spaces, and typically admit a two-parameter family of invariant metrics (this is proven in Appendix~\ref{orthoalgapp}). One of the parameters is the overall scale, and there is one additional essential parameter, which drops out only in the symmetric space case.

In the GN-type models in question, though, there are no free parameters, apart from the coupling constant (which corresponds to the overall scale). Thus, a reasonable question is what metric out of the whole family of metrics is realized in these models. As could be anticipated on general grounds~\cite{BykovComplex}, the resulting metric is the so-called \emph{normal} (reductive) metric on the homogeneous space. It is not K\"ahler unless the space is symmetric\footnote{For the $\tSU(n)$ case this was proven in~\cite{BykovGauged}, whereas in the present $\tO(n)$ or $\tSp(2n)$ cases the proof is given in Appendix~\ref{orthoalgapp}).}.

In this section we will take a direct approach, passing from the GN form to the geometric form of the sigma model, which allows reading off the metric. To this end, we add interactions to the free system discussed earlier:
\bea\label{LagrOrthoCompl}
\mathcal{L}=\mathrm{Tr}\left( \thickbar{D}W\omega_{2m}W^t \mathds{h}_n\right)+\mathrm{Tr}\left( \thickbar{W}^t\omega_{2m}D\thickbar{W} \mathds{h}_n\right)+{\vkappa\over 4\pi}\, \mathrm{Tr}(J_{\mathsf{o}} \thickbar{J}_{\mathsf{o}})\,,\qquad \frac{1}{2\pi}J_{\mathsf{o}}:=\mu_{\mathsf{o}}\,.
\eea
Here $\vkappa$ is the coupling constant, and interactions are constructed using the moment map~(\ref{SOmommap2}). It is convenient to rewrite this Lagrangian in terms of the $U,~V$ fields:
\bea\label{LagrOrthoCompl2}
\mathcal{L}\simeq 2\,\mathrm{Tr}\left( V\mathds{h}_n \thickbar{D}U+\thickbar{U}\mathds{h}_n D\thickbar{V}\right)+\mathrm{Tr}\left(\mathcal{A}_+U^t \mathds{h}_nU-\thickbar{\mathcal{A}_+}\thickbar{U} \mathds{h}_n\thickbar{U}^t\right)+{\vkappa\over 4\pi}\, \mathrm{Tr}(J_{\mathsf{o}} \thickbar{J}_{\mathsf{o}})\,.
\eea
Here the covariant derivatives acting on the  $U$ 
and $V$ fields are  $\thickbar{D}U=\thickbar{\partial} U- U\thickbar{\mathcal{A}}$ and $\thickbar{D}V=\thickbar{\partial} V- V\thickbar{\mathcal{A}}$, with similar expressions for the conjugate variables.

As discussed above, the gauge group is $\tGL(m, \CC) \ltimes \mathrm{Mat}_{m}^{\mathrm{symm}}(\CC)$, and gauge transformations act as follows:
\bea
\label{PartialGaugeOGr}
U \mapsto U g\,,\quad\quad V\mapsto g^{-1}V+q\, U^t,\quad\quad g\in \tGL(m, \CC)\,,\quad q\in \mathrm{Mat}_{m}^{\mathrm{symm}}(\CC)\,.
\eea
Notice, in particular, that the moment map~(\ref{SOmommap2}) is invariant w.r.t. these transformations. As a result, the $(V, \thickbar{V})$ Hermitian form in the interaction term of~(\ref{LagrOrthoCompl}) is degenerate\footnote{For example, if $m=1$, the interaction term can be written in the form
\begin{equation}
\mathcal{L}_{\mathrm{int}}=V\Big(|U|^2\,\mathds{1}_{n}-\thickbar{U}\otimes U\Big)\thickbar{V},
\end{equation}
so that it is proportional to the unitary projector. It means that the problem of deriving the sigma model metric is completely analogous to the problem of finding the QED propagator.}. Thus, prior to integrating over $V$ and $\thickbar{V}$, one needs to pick a gauge. A suitable partial gauge is
\bea
\thickbar{U}U=\mathds{1}_m,\quad\quad \frac{1}{2\pi}\mathcal{F}:=\thickbar{V}^tU+U^t \thickbar{V}=0\,.
\eea
This reduces the gauge freedom down to $\tU(m)\subset \tGL(m, \CC)$. For calculational purposes it is in fact simplest to use the generalized `$\alpha$-gauge', which amounts to adding a term ${1\over 2\alpha}\,\mathrm{Tr}(\mathcal{F} \thickbar{\mathcal{F}})$ to the Lagrangian\footnote{Formally this can be achieved by inserting $\delta(\mathcal{F}-M) \delta(\thickbar{\mathcal{F}}-\thickbar{M})$ in the path integral and then integrating over $M, \thickbar{M}$ with the weight $e^{-{1\over \alpha}\mathrm{Tr}(M\thickbar{M})}$. This is a direct analogue of $\alpha$-gauge in non-Abelian gauge theory, where one adds the term ${1\over \alpha}\,\mathrm{Tr}(\dd_\mu A_{\mu})^2$ to the gauge-invariant Lagrangian.}. The case $\alpha\to 0$ would then correspond to the gauge above, however it is more convenient to set ${\alpha^{-1}}={\vkappa\over 2\pi}$, in which case the interaction term in the gauge fixed Lagrangian acquires the form (here we use $\mathds{h}_n\thickbar{\mathds{h}}_n=\mathds{1}_n$)
\bear\label{SOgaugefixed}
{\vkappa\over 4\pi}\, \left[\mathrm{Tr}(J_{\mathsf{o}} \thickbar{J}_{\mathsf{o}})+\mathrm{Tr}(\mathcal{F} \thickbar{\mathcal{F}})\right]={2\pi\vkappa} \,\mathrm{Tr}\left[\thickbar{V}V\left(\mathds{1}_{n}+\thickbar{U}^t U^t\right)\right].
\eear
As a result, we have effectively diagonalized the quadratic form standing in the interaction term. Integrating over $V$ and $\thickbar{V}$, we get the sigma model Lagrangian in geometric form:
\bear \label{metrVbarVintegr}
\mathcal{L}_{\mathrm{GF}}={4\over 2\pi\vkappa}\,\mathrm{Tr}\left[D\thickbar{U}\, \mathds{h}_n\,\frac{1}{\mathds{1}_{n}+\left(U \thickbar{U}\right)^t}\,\mathds{h}_n\,\thickbar{D}U\right]+\mathrm{Tr}\left(\mathcal{A}_+ U^t \mathds{h}_n U-\thickbar{\mathcal{A}_+}\thickbar{U} \mathds{h}_n\thickbar{U}^t\right)=\\ \nonumber
={4\over 2\pi\vkappa}\,\mathrm{Tr}\left[D\thickbar{U}\,\mathds{h}_n\left(\mathds{1}_{n}-{1\over 2}\left(U \thickbar{U}\right)^t\right)\mathds{h}_n\,\thickbar{D}U\right]+\mathrm{Tr}\left(\mathcal{A}_+ U^t \mathds{h}_nU-\thickbar{\mathcal{A}_+}\thickbar{U} \mathds{h}_n\thickbar{U}^t\right)\,.
\label{MetricOGr}
\eear
When inverting the matrix, we used the normalization condition  $\thickbar{U}U=\mathds{1}_{m}$. Due to the isotropy constraint $U^t \mathds{h}_n U=0$, we can rewrite the above Lagrangian as
\bear\label{orthomnlagr}
&&\mathcal{L}_{\mathrm{GF}}={4\over 2\pi\vkappa}\,\mathrm{Tr}\left[D\thickbar{U}\,\thickbar{D}U\right]-{1\over \pi\vkappa}\mathrm{Tr}\left[\dd\thickbar{U}\,\mathds{h}_n\left(U \thickbar{U}\right)^t\mathds{h}_n\,\thickbar{\dd}U\right]+\\ \nonumber &&\quad\quad\quad +\;\mathrm{Tr}\left(\mathcal{A}_+ U^t \mathds{h}_n U-\thickbar{\mathcal{A}_+}\thickbar{U} \mathds{h}_n\thickbar{U}^t\right)\,.
\eear
The first term coincides with an analogous term for the Grassmannian $\mathsf{Gr}(m,n)$. In particular, the $\tGL(m, \CC)$ gauge field $\mathcal{A}$ only enters in this term, and its elimination results in the standard (Fubini-Study) metric on the Grassmannian. The second term represents a correction to it, and, due to the third term, one has to restrict to ${U^t\mathds{h}_n U=0}$. Since the constraint is holomorphic, the induced metric remains Hermitian.

As one can see from~(\ref{orthomnlagr}), the fundamental Hermitian form is
\bea\label{orthoHermform}
\Omega=\Omega^{(FS)}-{i\over 2}\,\mathrm{Tr}\left[d\thickbar{U}\, \mathds{h}_n \left(U \thickbar{U}\right)^t \mathds{h}_n\wedge dU\right]\,,
\eea
where
\bea
\Omega^{(FS)}=i\,\mathrm{Tr}\left[d\thickbar{U} \left(\mathds{1}_n-U \thickbar{U}\right)\wedge dU\right]\,.
\eea
is the Fubini-Study form on the Grassmannian $\mathsf{Gr}(m,n)$.

\subsubsection{The normal metric.}\label{orthoreductmetr}

We will now prove that the sigma model metric featuring in~(\ref{orthomnlagr}) is the normal metric. For a homogeneous space $G/H$, the latter is defined as follows. Let $\mathsf{g}=\mathsf{h}\oplus \mathsf{m}$ be the Lie algebra decomposition (here $\mathsf{m}$ is the orthogonal complement to $\mathsf{h}$ w.r.t. the Killing metric), and $J=-g^{-1}dg=J_{\mathsf{h}}+J_{\mathsf{m}}$ the corresponding decomposition of the Maurer-Cartan one-form. Then the normal metric is
\bea\label{reductmetr}
\left(ds^2\right)_{\textrm{normal}}=-\mathrm{Tr}\left(J_{\mathsf{m}}^2\right)\,.
\eea
To compute $J_{\mathsf{m}}$, it will be somewhat easier to work with the non-diagonal form of $\mathds{h}_n$:
\bea\label{symmformsplit}
\mathds{h}_n=\begin{pmatrix}
    0 & \mathds{1}_m & 0\\
    \mathds{1}_m & 0 & 0\\
    0 & 0 & \mathds{1}_{n-2m}
\end{pmatrix}\,.
\eea
Given the matrix $U$ of $m$ isotropic orthonormalized vectors, satisfying $U^t \mathds{h}_n U=0$ and $\thickbar{U}U=\mathds{1}_m$, construct the group element
\bea\label{Ogroupelem}
g=\left(U\; \mathds{h}_n U^\ast\;Y\right)
\,,
\eea   
where $Y$ is the matrix of $n-2m$ complimentary orthonormal vectors, satisfying ${\thickbar{Y}U=0},\, Y^t \mathds{h}_n U=0$. 
These constraints follow from the requirement that $g$ be orthogonal, $g \in \tO(n)$. For more on this parametrization see Appendix~\ref{OGrHomSpacesec}.

The advantage of working with the non-diagonal form~(\ref{symmformsplit}) of $\mathds{h}_n$ is that the space $\mathsf{m}$ is particularly easy to describe:
\bea
\mathsf{m} =\begin{pmatrix}
    \mathbf{0}_m & \bullet & \bullet\\
   \bullet & \mathbf{0}_m & \bullet\\
    \bullet & \bullet & \mathbf{0}_{n-2m}
    \end{pmatrix}\,.
\eea
Here $\mathbf{0}_m$ is zero-block of size $m\times m$. 
As a result, we compute
\bea
J_{\mathsf{m}}=\left[-g^{-1}dg\right]_{\mathsf{m}}=-\begin{pmatrix}
    \mathbf{0}_m & \thickbar{U}\mathds{h}_n dU^\ast & \thickbar{U} dY\\
   U^t \mathds{h}_n dU & \mathbf{0}_m & U^t \mathds{h}_n dY\\
    \thickbar{Y}dU & \thickbar{Y} \mathds{h}_n dU^\ast & \mathbf{0}_{n-2m}
\end{pmatrix}\,.
\eea
The metric~(\ref{reductmetr}) is then
\bear
&&\left(ds^2\right)_{\textrm{normal}}=
4\,\mathrm{Tr}\left(d\thickbar{U} \left(1-U\thickbar{U}\right) dU\right)-2\,\mathrm{Tr}\left(d\thickbar{U}\,\mathds{h}_n \left(U \thickbar{U}\right)^t \mathds{h}_n dU \right) \,,\quad\quad
\eear
where we have used the completeness relation $U\thickbar{U}+\mathds{h}_n(U\thickbar{U})^t \mathds{h}_n+Y\thickbar{Y}=\mathds{1}_n$ to eliminate~$Y, \thickbar{Y}$. Clearly, this expression coincides with the metric in~(\ref{orthomnlagr}).

\vspace{0.3cm}
In Appendix~\ref{invmetrapp} we prove that the normal metric is K\"ahler in only two special cases, namely the ones of minimal ($m=1$) and maximal ($n=2m$) orthogonal Grassmannians. One way to understand this is as follows. All orthogonal Grassmannians admit K\"ahler metrics (cf.~\cite{SharpeExotic}), however in general such metrics are special points in the moduli space of invariant metrics, and do not coincide with the normal metric arising from the GN setup. The two extreme cases $m=1$ and $n=2m$ correspond to symmetric spaces. If a K\"ahler manifold is a symmetric space, the invariant metric is unique (up to overall scale), so that it coincides with the normal metric and is K\"ahler. Let us now consider these special cases in detail. 

\subsubsection{Minimal Grassmannian $m=1$.}
In this simplest case the fields $U$ and $\thickbar{U}$ are just row and column vectors, and this leads to a significant simplification. Indeed, the second term in (\ref{orthomnlagr}) vanishes, since it  is proportional to derivatives of the isotropy constraints\footnote{Alternatively, it can be eliminated by a shift of the gauge field $\mathcal{A}_+$.} $U^t\mathds{h}_n U=\thickbar{U}\mathds{h}_n\thickbar{U}^t=0$.  The resulting Lagrangian is
\bea\label{OGr1nLagr}
\mathcal{L}_{\mathsf{OGr}(1,n)}={4\over 2\pi\vkappa}\,D\thickbar{U}\,\thickbar{D}U\,.
\eea
It turns out that the metric is just the Fubini-Study metric of $\mathsf{CP}^{n-1}$ restricted to the  surface of constraints, thus it is obviously K\"ahler. The manifold itself is a quadric in $\mathsf{CP}^{n-1}$: when $n=3$, it is $\CP^1$ embedded into $\CP^2$ by means of the Veronese map, whereas when $n=4$, it is known to be isomorphic to $\mathsf{CP}^1\times\mathsf{CP}^1$ (cf.~\cite[Lecture 22]{Harris}).

\subsubsection{Maximal Grassmannian $n=2m$.} \label{maxOGrass} In this special case, besides the $m$ orthonormal vectors  $u_1, \ldots, u_m$ (columns of the matrix $U$, satisfying $\thickbar{U}U=\mathds{1}_m$) let us consider the additional vectors $\widetilde{u}_1=\mathds{h}_{2m}u_1^\ast, \ldots, \widetilde{u}_m=\mathds{h}_{2m}u_m^\ast$. These latter vectors are again orthonormal and, moreover, orthogonal to the ones of the first group, due to $U^t \mathds{h}_{2m}U=\widetilde{U}^\dagger U=0$. As a result, we obtain a complete basis of orthonormal vectors $u_1, \ldots, u_m, \mathds{h}_{2m}u_1^\ast, \ldots, \mathds{h}_{2m}u_m^\ast$ in $\CC^{2m}$. These vectors satisfy the partition of unity: $\sum_{i=1}^m (u_i\otimes \widebar{u_i}+\mathds{h}_{2m}u_i^\ast\otimes \widebar{u_i}^\ast \mathds{h}_{2m})=\mathds{1}_{2m}$, which in matrix terms may be written as
\bea
\mathds{h}_{2m}(U\thickbar{U})^t \mathds{h}_{2m}=\mathds{1}_{2m}-U\thickbar{U}\,.
\eea
Substituting in~(\ref{orthoHermform}), we find that in this case
\bea\label{maxOGrassform}
\Omega={1\over 2}\,\Omega^{(FS)}\,,\quad\quad\quad (n=2m)\,.
\eea
In particular, $d\Omega=0$, so that the metric is K\"ahler.

\section{Symplectic Grassmannians $\mathsf{SGr}(m, 2n)$}
\label{symplecticgrass}
In this section we apply the logic  proposed in the previous sections to the symplectic case. The structure of the construction is the same, the differences being mostly of technical nature.

The definition of symplectic Grassmannian is parallel to the one of orthogonal Grassmannian. Consider the vector space $\CC^{2n}$ with a non-degenerate \emph{skew}-symmetric tensor $\omega_{2n}$ on it. $\mathsf{SGr}(m, 2n)$ may be defined as the set of $m$-planes, isotropic w.r.t. $\omega_{2n}$. As before, we package the $m$ vectors in a single $m\times 2n$ matrix $U$, so that the isotropy constraint takes the form $U^t \omega_{2n}U=0$. Thus, $\mathsf{SGr}(m, 2n)$ is naturally embedded in the standard Grassmannian $\mathsf{Gr}(m, 2n)$.

Again, for technical simplicity we will require that the matrix $\omega_{2n}$ is real, additionally satisfying  $\omega_{2n}\thickbar{\omega}_{2n}=\mathds{1}_{2n}$ (equivalently, that $\omega_{2n}^2=-\mathds{1}_{2n}$). Otherwise the matrix $\omega_{2n}$ may be chosen at our will.

\subsection{Symplectic Grassmannians.}

Just as in the orthogonal case, one starts with pairs of canonical variables $U, V$, both of which are $m\times 2n$ matrices, i.e.
\bea
U\in \mathrm{Hom}(\CC^m, \CC^{2n}),\quad\quad V\in  \mathrm{Hom}(\CC^{2n}, \CC^m)\,.
\eea
The difference is that, in place of a symmetric matrix $\mathds{h}_n$ that one introduced in the case of orthogonal Grassmannians, here one has a skew-symmetric non-degenerate
tensor $\omega_{2n}$  on $\CC^{2n}$. One can think of it as a symplectic form with constant coefficients. Such form can only exist on even-dimensional spaces, which is the reason why we restrict to even-dimensional ambient space $\CC^{2n}$ in this section.

Next, we require that the columns of $U$ form a basis in an $m$-plane, isotropic w.r.t.~$\omega_{2n}$~\cite{SharpeExotic}:
\bear \label{symplcons}
&& H_{\mathrm{asymm}}:=U^t \omega_{2n} U=0\,,
\eear

Here  $H_{\mathrm{asymm}}$ is  an  $m\times m$ skew-symmetric matrix, $H_{\mathrm{asymm}}^t=-H_{\mathrm{asymm}}$, 
transforming homogeneously under the action of $\tGL(m, \CC)$, i.e. $H_{\mathrm{asymm}}\to g^t H_{\mathrm{asymm}} g$. Therefore the  group of symplectic reduction\footnote{Its  matrix realization is
\bear
&&\begin{pmatrix}
g&0\\
0& (g^{-1})^t
\end{pmatrix}\begin{pmatrix}
1&\beta\\
0& 1
\end{pmatrix}=\begin{pmatrix}
g&g\cdot\beta\\
0& (g^{-1})^t
\end{pmatrix},\quad \\ &&\textrm{with}\quad g\in \tGL(m, \CC),\quad \beta^t=-\beta\,.
\eear} is
\bea
\mathcal{G}=\tGL(m, \CC) \ltimes \mathrm{Mat}_{m}^{\mathrm{asymm}}(\CC)\,,
\eea
where the second factor is the space of anti-symmetric matrices, viewed as an additive abelian group. The dimension of the complex space of fields obtained after symplectic reduction is
\bea\label{symplgrdim}
\mathrm{dim}_{\CC}\Phi=4mn-2\,\mathrm{dim}_{\CC}\,\mathcal{G}=2\left(2mn-{3m^2-m\over 2}\right)=2\,\mathrm{dim}_{\CC}\mathsf{SGr}(m, 2n)\,,
\eea
which is twice the dimension of the symplectic Grassmannian. 
The concrete Gross-Neveu Lagrangian  for this case (more exactly, its holomorphic piece)  may be written in parallel to~(\ref{orthoGrassLagr}):
\bear\label{symplGrassLagr}
&& \mathcal{L}=\mathrm{Tr}\left( \thickbar{D}W \mathds{h}_{2m} W^t\omega_{2n}\right),\quad \\
\nonumber && \textrm{where}\quad\quad W=\begin{pmatrix}
U&V^t
\end{pmatrix},
\quad
\thickbar{D}W:=\thickbar{\dd}W-W\,\begin{pmatrix}
\thickbar{\mathcal{A}}&\thickbar{\mathcal{A}_+}\\
0& -\thickbar{\mathcal{A}}^t
\end{pmatrix},
\eear
where now $\thickbar{\mathcal{A}_+}$ is a skew-symmetric $m\times m$-matrix representing an abelian $\mathrm{Mat}_{m}^{\mathrm{asymm}}(\CC)$  gauge field and $\mathds{h}_{2m}=\begin{pmatrix}
0&\mathds{1}_m\\
\mathds{1}_m & 0
\end{pmatrix}$. Writing this out in components,
\bea\label{symplfree}
\mathcal{L}\simeq 2\,\mathrm{Tr}\left(V\omega_{2n} \thickbar{D}U\right)+
\mathrm{Tr}\left(\thickbar{\mathcal{A}_+} U^t\omega_{2n} U\right)\,.
\eea
Gauge transformations act as follows:
\bea\label{symplgaugetrans}
U \mapsto U g\,,\quad\quad V\mapsto g^{-1}V+q\, U^t,\quad\quad g\in \tGL(m, \CC)\,,\quad q\in \mathrm{Mat}_{m}^{\mathrm{asymm}}(\CC)\,.
\eea

We pass over to the description of interactions. Here the relevant moment map is
\begin{empheq}[box = \fcolorbox{black}{white}]{align}
\label{MomentMapSp}
\quad \mu_{\mathsf{sp}}=\left(U V+V^t U^t\right)\omega_{2n}\,. \quad
\end{empheq}
It is also instructive to compare this expression with the one we had for the orthogonal group, see~(\ref{SOmommap2}). 
One easily checks that $\mu_{\mathsf{sp}}^t\omega_{2n}+\omega_{2n} \mu_{\mathsf{sp}}=0$, so that the moment map indeed belongs to the symplectic Lie algebra:
$\mu_{\mathsf{sp}} \in \mathsf{sp}(2n)
$. 
Notice that the moment map is invariant w.r.t. the gauge transformations~(\ref{symplgaugetrans}). Besides, it is equivariant w.r.t. the $\tSp(2n, \CC)$ transformations
\bea\label{globtrans}
U\to g_0 U,\quad\quad V\to V g_0^t,\quad\quad g_0\in \tSp(2n, \CC)\,.
\eea
Indeed, since $g_0^t\omega_{2n} g_0=\omega_{2n}$, one finds 
$
\mu_{\mathsf{sp}}\to g_0 \mu_{\mathsf{sp}} g_0^{-1}\,.
$ Besides, just like in the unitary and orthogonal cases, we find that $\mu_{\mathsf{sp}}$ is nilpotent:
\bea\label{munilpeq3}
\mu_{\mathsf{sp}}^2=UV\omega_{2n}V^t U^t \omega_{2n}\,,\quad\quad \textrm{and}\quad\quad \mu_{\mathsf{sp}}^3=0\,.
\eea
To arrive at this, one uses the constraints $V\omega_{2n}U=0$ and $U^t \omega_{2n} U=0$ that follow by varying the Lagrangian~(\ref{symplGrassLagr}) w.r.t. the gauge fields.

\subsection{Sigma model metrics on symplectic Grassmannians.}

In order to deduce the geometric form of the sigma model from the GN formulation, first of all we add to the free Lagrangian the interaction piece
\bea
\label{InteractionSp}
\mathcal{L}_{\mathrm{int}}={\vkappa\over 4\pi}\,\mathrm{Tr}\left(J_{\mathsf{sp}} \thickbar{J_{\mathsf{sp}}}\right),\qquad\frac{1}{2\pi}J_{\mathsf{sp}}:=
\mu_{\mathsf{sp}}\,.
\eea
This is only invariant w.r.t. to a subset of the transformations~(\ref{globtrans}), where the matrix $g_0$ additionally satisfies $\thickbar{g_0}g_0=\mathds{1}_{2n}$, so that the symmetry group is reduced to the compact subgroup $\mathbf{Sp}(2n)$. This  is the true symmetry group for sigma models of symplectic Grassmannians.

Just like in the orthogonal case, one has to pick a gauge for the gauge symmetry~(\ref{symplgaugetrans}). A suitable gauge is
\bea
\thickbar{U}U=\mathds{1}_m,\quad\quad \frac{1}{2\pi}\mathcal{F}:=\thickbar{V}^t U-U^t \thickbar{V}=0\,.
\eea
In order to promote it to the $\alpha$-gauge, we add to the gauge-invariant interacting Lagrangian the term ${1\over 2\alpha}\,\mathrm{Tr}(\mathcal{F} \thickbar{\mathcal{F}})$. Just as before, it is convenient to set ${\alpha^{-1}}={\vkappa\over 2\pi}$, in which case the Lagrangian acquires the form
\bear
&\mathcal{L}_{\mathrm{GF}}=2\,\mathrm{Tr}\left(V\omega_{2n}\thickbar{D}U+\thickbar{U}\omega_{2n} D\thickbar{V}\right)+{\vkappa\over 4\pi}\, \left[\mathrm{Tr}(J_{\mathsf{sp}} \thickbar{J}_{\mathsf{sp}})+\mathrm{Tr}(\mathcal{F} \thickbar{\mathcal{F}})\right]=\\ \nonumber
&=2\,\mathrm{Tr}\left(V\omega_{2n}\thickbar{D}U+\thickbar{U}\omega_{2n} D\thickbar{V}\right)+2\pi\vkappa \,\mathrm{Tr}\left[\thickbar{V}V\left(\mathds{1}_{n}+\thickbar{U}^t U^t\right)\right]\,.
\eear
For brevity we are not writing out explicitly the second term in~(\ref{symplfree}), but one should  remember that the $\thickbar{\mathcal{A}_+}$ gauge field imposes the constraint $U^t \omega_{2n} U=0$. Integration over $V, \thickbar{V}$ is now straightforward and leads to the following, in full analogy with~(\ref{metrVbarVintegr}):
\bea
\mathcal{L}_{\mathrm{GF}}={4\over 2\pi\vkappa}\,\mathrm{Tr}\left[D\thickbar{U}\,\omega_{2n}^{-1} \left(\mathds{1}_{n}-{1\over 2}\left(U \thickbar{U}\right)^t\right)\omega_{2n}\,\thickbar{D}U\right]\,,
\eea
where we have used $\omega_{2n}^{-1}=-\omega_{2n}$.  
Using the isotropy constraint, one can again simplify the second term, arriving at 
\bea\label{lagrsympl2}
\mathcal{L}_{\mathrm{GF}}={4\over  2\pi\vkappa}\,\mathrm{Tr}\left[D\thickbar{U}\,\thickbar{D}U\right]-{1\over \pi\vkappa}\mathrm{Tr}\left[\dd\thickbar{U}\,\omega_{2n}^{-1}\left(U \thickbar{U}\right)^t \omega_{2n}\,\thickbar{\dd}U\right]\,. 
\eea
The first term leads to the Fubini-Study metric on $\mathsf{Gr}(m, 2n)$, and the second term is a correction.

\subsubsection{The normal metric.} As in the orthogonal case, we can prove that the metric in~(\ref{lagrsympl2}) is normal, see section~\ref{orthoreductmetr} for the definition. Given the $m\times n$ matrix $U$ satisfying $U^t \omega_{2n} U=0$ and $\thickbar{U}U=\mathds{1}_m$, we construct the group element $g=(U\;\omega_{2n}^{-1} U^\ast\; Y)$, a direct generalization of~(\ref{Ogroupelem}) in the orthogonal case. We describe this parametrization in detail in Appendix~\ref{SGrHomSpacesec}.

Next, we need to compute the $\mathsf{m}$-component of the Maurer-Cartan current $J_{\mathsf{m}}$.  It will be useful to write the  symplectic form as a $3\times 3$ block matrix:
\bea\label{omegasymplsplit}
\omega_{2n}=\begin{pmatrix}
    0 & \mathds{1}_m & 0\\
    -\mathds{1}_m & 0 & 0 \\
    0 & 0 & \omega_{2n-2m}
\end{pmatrix}\,.
\eea
In this case $\mathsf{m}$ comprises the off-diagonal matrices:
\bea
\mathsf{m} =\begin{pmatrix}
    \mathbf{0}_m & \bullet & \bullet\\
   \bullet & \mathbf{0}_m & \bullet\\
    \bullet & \bullet & \mathbf{0}_{n-2m}
\end{pmatrix}\,.
\eea
Computing the $\mathsf{m}$-part of $J=-g^{-1}dg$ and then the metric $ds^2=-\mathrm{Tr}(J_{\mathsf{m}}^2)$, one finds agreement with the one in~(\ref{lagrsympl2}).

\subsubsection{The maximal Grassmannian.} \label{maxSGrass} Just as in the orthogonal case, the maximal Grassmannian $m=n$ is special in that the resulting metric is K\"ahler\footnote{However, unlike the orthogonal case, here $m=1$ is not in any way special.}. The proof is parallel to the one of the orthogonal case in section~\ref{maxOGrass}.

Here, besides the $m$ orthonormal vectors  $u_1, \ldots, u_m$ (columns of the matrix $U$, satisfying $\thickbar{U}U=\mathds{1}_m$) we consider the additional vectors $\widetilde{u}_1=\omega_{2m} u_1^\ast, \ldots, \widetilde{u}_m=\omega_{2m} u_m^\ast$. These latter vectors are again orthonormal and orthogonal to the ones of the first group, due to $U^t\omega_{2m} U=-\widetilde{U}^\dagger U=0$. As a result, we obtain a complete basis of orthonormal vectors $u_1, \ldots, u_m, \omega_{2m} u_1^\ast, \ldots, \omega_{2m} u_m^\ast$ in $\CC^{2m}$. These vectors satisfy the partition of unity: $\sum_{i=1}^m (u_i\otimes \thickbar{u}_i+\widetilde{u}_i^\ast\otimes \thickbar{\widetilde{u}_i})=\mathds{1}_{2m}$, which in matrix terms may be written as
\bea
\omega_{2m} (U\thickbar{U})^t \omega_{2m}^{-1}=\mathds{1}_{2m}-U\thickbar{U}\,.
\eea
Substituting in~(\ref{lagrsympl2}), we find that in this case the K\"ahler form is
\bea\label{SGrassmaxform}
\Omega={1\over 2}\,\Omega^{(FS)}\,,\quad\quad\quad n=m\,.
\eea
In particular, $d\Omega=0$, so that the metric is K\"ahler.

\section{Elimination of the gauge fields}
\label{gaugeAzero}

All of the models that we have discussed so far schematically have the following Lagrangian\footnote{To simplify the expressions, here we omit the terms with the gauge fields $\mathcal{A}_+$ and $\thickbar{\mathcal{A}_+}$.}:
\bea\label{lagrgen}
\mathcal{L}=2\,\mathrm{Tr}\left(V\thickbar{D}U+\thickbar{U}D\thickbar{V}\right)+{2\pi\vkappa}\left(\text{interactions}\right)\,.
\eea
To study the $\beta$-function of such models, we first need to decide what to do with the gauge fields. One option is to impose a gauge constraint on the matter fields $U, V$ and to explicitly resolve the constraints that couple to the gauge fields $\mathcal{A}$ and $\thickbar{\mathcal{A}}$. For the $\CP^{n-1}$-model,  choosing inhomogeneous coordinates (i.e. setting $U_n=1$) results in such gauge, for example. The drawback is that the full global symmetry is no longer manifest.

A better option is to impose a gauge condition on the gauge field itself. Remarkably, the best gauge is to simply set it to zero:
\bea\label{AbarAzero}
\mathcal{A}=\thickbar{\mathcal{A}}=0\,.
\eea
Clearly, this is not a typical gauge condition that one encounters in gauge theory. The reason it is admissible in the present setup is that in our models the gauge field is essentially topological (see~\cite{BykovRiemann} for more details).

\subsection{Abelian case.}

In explaining why one can impose~(\ref{AbarAzero}), we will start with the abelian case. Here gauge transformations have the form
\bea
\thickbar{\mathcal{A}}\; \mapsto \;  
\thickbar{\mathcal{A}}+ \thickbar{\dd} \chi(z, \thickbar{z}\,) \,,
\eea
with $\chi$ a complex-valued function. By the Cauchy-Green formula, we may set the transformed gauge field to zero by choosing
\bea
\chi(z, \thickbar{z})={i\over \pi}\,\int d^2w\,\frac{\thickbar{\mathcal{A}}(w, \thickbar{w}\,)}{z-w}\,.
\eea
The integral converges, provided that $\thickbar{\mathcal{A}}=\mathcal{O}\left({1\over r^{1+\delta}}\right)$ for $\delta>0$ as $r=|w|\to\infty$. This asymptotic behavior also ensures that the first Chern class of the bundle vanishes:
\bea
\int_{\mathbb{R}^2} c_1={1\over 2\pi}\,\int_{\mathbb{R}^2} \;dA=\underset{r\to \infty}{\mathrm{lim}}\;{1\over 2\pi}\oint_{C_r} \;A=0\,,
\eea
where $C_r$ is a circle of radius $r$, and  $A=i\left(\mathcal{A}dz-\thickbar{\mathcal{A}}d\,\thickbar{z}\,\right).$

\subsection{Non-Abelian case.}

The non-Abelian case is more involved. Here the relevant gauge transformation is
\bea
\thickbar{\mathcal{A}}\; \mapsto \;  
\thickbar{\mathcal{A}}+ \thickbar{\dd} g \cdot g^{-1}\, \,,
\eea
where $g\in \mathrm{G}_{\CC}$ is a group element of the corresponding complex gauge group (in the abelian case above $\mathrm{G}_{\CC}=\CC^\ast$ and $g=e^{\chi}$).  We would thus like to solve the equation
\bea\label{gaugetranszero}
\thickbar{\dd} g+\thickbar{\mathcal{A}} \,g=0\,.
\eea
For generic $\thickbar{\mathcal{A}}$ one cannot solve it analytically. However, one can establish existence of a solution by perturbation theory (as was originally shown in~\cite{Nijenhuis, Rawnsley}; see also~\cite[Section 2.2]{DK}). Notice that under a rescaling $(z, \thickbar{z}\,) \to \upepsilon \cdot  (z, \thickbar{z}\,)$ one effectively has a rescaling
\bea
\thickbar{\mathcal{A}} \to \upepsilon\, \thickbar{\mathcal{A}}\,.
\eea
This means that we can make the absolute value of the `perturbation' $\big|\thickbar{\mathcal{A}}\,\big|$ as small as we want by working in a sufficiently small neighborhood of any given point $(z, \thickbar{z}\,)$ on the worldsheet. 
Consider a small disc of radius $\upepsilon$ around an arbitrary point on our Riemann surface. For simplicity we will assume this point is $z=\thickbar{z}=0$, so that the disc is $D_{\upepsilon}=\{\;|z|<\upepsilon\;\}$. Next, we write an integral equation for the function $g(z, \thickbar{z}\,)$ defined in this disc:
\bea\label{integreq2}
g(z, \thickbar{z}\,)=f(z)+{i\over \pi}\,\int_{D_{\upepsilon}} d^2w\,\frac{1}{z-w}\,\thickbar{\mathcal{A}} \,g(w, \thickbar{w}\,)\,,\quad\quad (z, \thickbar{z}\,)\in D_{\upepsilon}\,,
\eea
where $f(z)$ is a holomorphic matrix-valued function. 
A solution to this equation would automaticaly satisfy~(\ref{gaugetranszero}) inside the disc. Indeed, differentiating~(\ref{integreq2}) w.r.t. $\thickbar{z}$, one arrives at~(\ref{gaugetranszero}).

Picking $f(z)=\mathds{1}$, we may set up a perturbative expansion in $\thickbar{\mathcal{A}}$. Assuming $\big|\thickbar{\mathcal{A}}\,\big|<C$, we may estimate
\bea
\bigg|{i\over  
\pi}\,\int_{D_{\upepsilon}} d^2w\,\frac{1}{z-w}\,\thickbar{\mathcal{A}} \,g(w, \thickbar{w}\,)\bigg|< \mathrm{const}.\, \upepsilon \,C\,\big|g\big|\,,
\eea
ensuring convergence for sufficiently small $\upepsilon$ (note that the bound on $\upepsilon$ depends only on $C$). As a result, in an $\upepsilon$-neighborhood of any point we may set $\mathcal{A}=\thickbar{\mathcal{A}}=0$ by a gauge transformation $g(z, \thickbar{z}\,)$. In the intersection of two such neighborhoods $D_1 \cap D_2$ the two matrices are related by a \emph{holomorphic} gauge transformation
\bea
g_2(z,\thickbar{z}\,)=\mathbf{g}_{12}(z)\cdot g_1(z,\thickbar{z}\,)\,.
\eea
The set of such neighborhoods and holomorphic gauge transformations between them defines the structure of a holomorphic vector bundle over the Riemann surface. The field $U$ in~(\ref{lagrgen}) is then a section of this bundle, meaning that it undergoes a gauge transformation $U \mapsto \mathbf{g}_{12}(z) U$ in the overlap of two patches, whereas $V$ is a section of the dual bundle. If the bundle is trivial, then, by definition, we may choose all transition matrices to be trivial, that is $\mathbf{g}_{ij}(z)=\mathds{1}$, so that there exists a global gauge transformation $g(z, \thickbar{z}\,)$ setting $\mathcal{A}=\thickbar{\mathcal{A}}=0$. Henceforth we will assume that we are dealing with a trivial holomorphic bundle, so that this gauge may be chosen.

\section{One-loop $\beta$-function} \label{betafuncsec}

The goal of the present section is to calculate one-loop $\beta$-functions for the three families of Grassmannians~(\ref{UGr})-(\ref{SGr}). In the `geometric' approach to sigma models, where the action is formulated in terms of the metric and $B$-field, this is usually done with the help of the background field method (cf.~\cite{Ketov} and references therein). Here instead the calculation reduces to the analysis of certain elementary Feynman diagrams, akin to the ones of $\varphi^4$-theory.

Our result will be that, independent of the concrete model in question, the $\beta$-function is proportional to the dual Coxeter number\footnote{See~\cite{Kac} for the definition and~\cite{vanRitbergen1998} for a physics-oriented discussion.} $h^\vee$ of the respective symmetry algebra (corresponding to the isometry group of the homogeneous space). In the case of symmetric spaces, this is a well-known result, cf.~\cite{Zarembo} or~\cite[Chapter 15]{ZJ}.  For later use, we collect the values of $h^\vee$ for the relevant Lie algebras in the following table:

\begin{table}[H]
\begin{center}
\begin{tabular}{|c  c c |} 
 \hline
 Notation & Lie algebra & $h^\vee$  \\ [0.5ex] 
 \hline\hline
 $A_{n}$ & $\mathsf{sl}(n+1)$ & $n+1$ \\ 
 \hline
 $B_n$ & $\mathsf{o}(2n+1)$ & $2n-1$  \\
 \hline
 $C_n$ & $\mathsf{sp}(2n)$ & $n+1$ \\
 \hline
 $D_n$ & $\mathsf{o}(2n)$ & $2n-2$ \\ [1ex] 
 \hline
\end{tabular}
\caption{\label{CoxeterTable}Dual Coxeter numbers of classical Lie algebras.}
\end{center}
\end{table}

It is also useful to observe that, for an arbitrary orthogonal Lie algebra $\mathsf{so}(n)$, the dual Coxeter number is simply $h^\vee=n-2$.

\subsection{$\beta$-function in the $\mathsf{CP}^{n-1}$ case.}

Before passing to general Grassmannians,  it is helpful to discuss the simplest case $\mathsf{Gr}(1,n)\simeq\mathsf{CP}^{n-1}$. As we shall see, diagrammatic calculations for arbitrary Grassmannians are simple generalizations of the $\mathsf{CP}^{n-1}$ case.

\begin{figure}[h!]
\vspace{0.5cm}
\centering
\begin{overpic}[scale=0.7,unit=1mm]{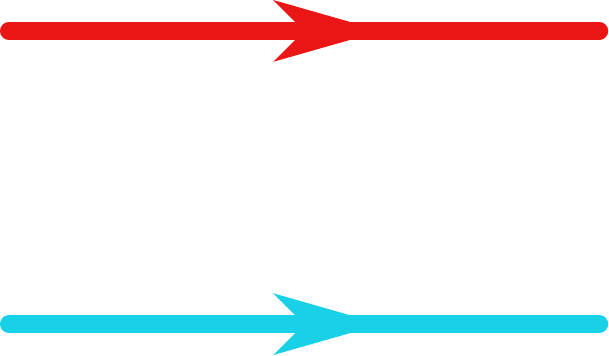}
\put(-43,22){$\Big\langle U_a(z,\thickbar{z}\,)V_b(w,\thickbar{w}\,)\Big\rangle_0=$}
\put(-43,1){$\Big\langle \thickbar{U}_a(z,\thickbar{z}\,)\thickbar{V}_b(w,\thickbar{w}\,)\Big\rangle_0=$}
\put(45,22){$=\dfrac{\delta_{ab}}{2\pi(z-w)}$\,,}
\put(45,1){$=\dfrac{\delta_{ab}}{2\pi(\thickbar{w}-\thickbar{z}\,)}$\,.}
\put(0,25){$a$}
\put(41,25){$b$}
\put(0,4){$a$}
\put(41,4){$b$}
\end{overpic}
\caption[Fig1-1]{Elementary Green's functions (`propagators'). The index $0$ denotes  averaging w.r.t. the free action and Latin indices denote vector components. Henceforth we assume that Latin indices run from $1$ to $n$.}
\label{Propagators}
\vspace{0.2cm}
\end{figure}

\begin{figure}[h!]
\centering
\begin{overpic}[scale=0.7,unit=1mm]{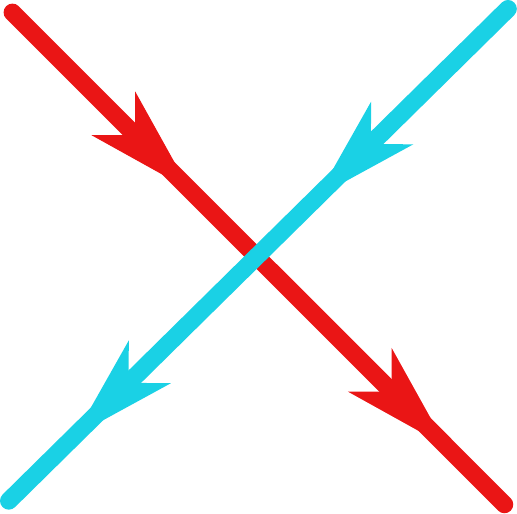}
\end{overpic}
\caption[Fig1-1]{The vertex.}
\label{VertexUGr}
\vspace{1cm}
\end{figure}

The corresponding Gross-Neveu Lagrangian, with the gauge field set to zero,  is
\bea
\label{LagrangianGNgaugedCP}
\mathcal{L}=2\left(V\thickbar\partial U+\thickbar{U}\partial\thickbar{V}\right)+\frac{\vkappa}{2\pi}\,\mathrm{Tr}\big(J_{\mathsf{cp}}\thickbar{J}_{\mathsf{cp}}\big),\qquad\frac{1}{2\pi}J_{\mathsf{cp}}:=U\otimes V\,.
\eea
Here the $U$ and $V$ fields are row and column vectors, and elementary Green's functions of these fields are given in  Figure~\ref{Propagators}.
\vspace{0.3cm}
These are the so-called symplectic bosons in the language of conformal field theory~\cite{Goddard}. To clarify the calculations of Feynman diagrams we rewrite the interaction in terms of field components:
\begin{empheq}[box = \fcolorbox{black}{white}]{align}
\label{InteractionGrm}
\quad
\mathcal{L}_{int}=\frac{\vkappa}{2\pi}\,\mathrm{Tr}\big(J_{\mathsf{cp}}\thickbar{J}_{\mathsf{cp}}\big)=2\pi\vkappa\left(U_{a}V_{b}\thickbar{V}_{b}\thickbar{U}_{a}\right).\quad 
\end{empheq}
The relevant vertex is drawn in Fig.~\ref{VertexUGr}. 

\begin{figure}[h!]
\centering
\begin{overpic}[scale=0.5,unit=1mm]{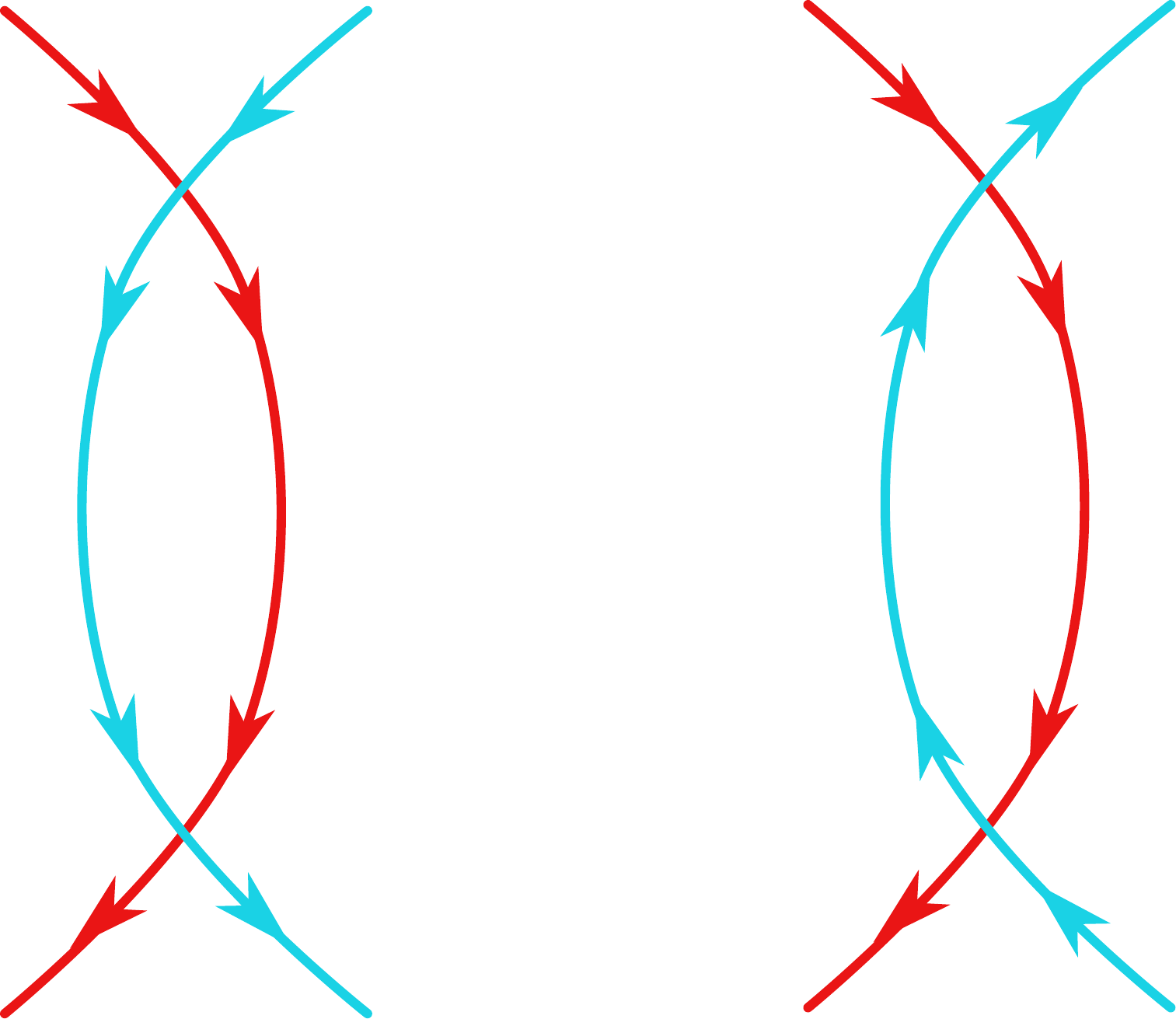}
\put(-5,65){$U_a$}
\put(25,65){$\thickbar{U}_b$}
\put(-5,-0.5){$V_c$}
\put(25,-0.5){$\thickbar{V}_d$}
\put(48,65){$U_a$}
\put(48,-0.5){$V_c$}
\put(77.5,65){$\thickbar{V}_d$}
\put(78,-0.5){$\thickbar{U}_b$}
\end{overpic}
\caption[Fig1-1]{Two types of divergent diagrams.}
\label{DivDiagrams}
\end{figure}

To calculate the $\beta$-function of the model one needs to consider a four-point Green's function of the type $\big\langle\,U_a\,\thickbar{U}_b\,V_c\,\thickbar{V}_d\,\big\rangle.$ At one-loop level there are two types of divergent diagrams (see Fig.~\ref{DivDiagrams}). It is not hard to see~\cite{BykovBetaFunction} that each of these diagrams is proportional to the following elementary integral\footnote{We set two of the four external momenta in the four-point Green's function to zero. By momentum conservation, the diagrams then depend on a single momentum $p$.}:
\bea
\frac{I(p)}{2\pi}=\int \frac{d^2 z_{12}}{2\pi}\,e^{i(p,z_{12})}\,\frac{1}{|z_{12}|^2}=-\frac{1}{2}\ln\bigg(\frac{p^2}{\Lambda^2}\bigg)+\mathrm{const}\,,
\eea
where $p$ is the external momentum, $z_{12}=z_1-z_2$ and $\Lambda$ is the UV cutoff. Note that the diagrams have a symmetry factor of $2$ and also differ in sign due to the opposite directions of the blue line.


\begin{figure}[h!]
\centering
\begin{overpic}[scale=0.7,unit=1mm]{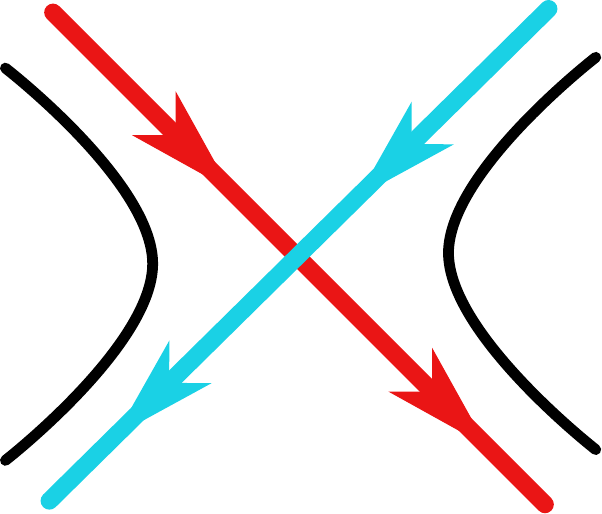}
\end{overpic}
\caption[Fig1-1]{The modified vertex.}
\label{VertexUGr1}
\end{figure}

\vspace{-0.7cm}
Thus, we reduce the one-loop calculations to the problem of determining the tensor structure of the divergent diagrams. To this end, we modify the Feynman rules in order to visualize the tensor structure of the interaction: the resulting  vertex is shown in Figure \ref{VertexUGr1}.
Here black lines denote contractions of the flavor indices,  similarly to the theory of the $1/n$ expansion \cite{Vasiljev2,ZJ,Makeenko}. In this notation a closed line corresponds to the trace of a unit flavor matrix, thus giving a factor of $n$.
\begin{figure}[h!]
\centering
\begin{overpic}[scale=0.7,unit=0.8mm]{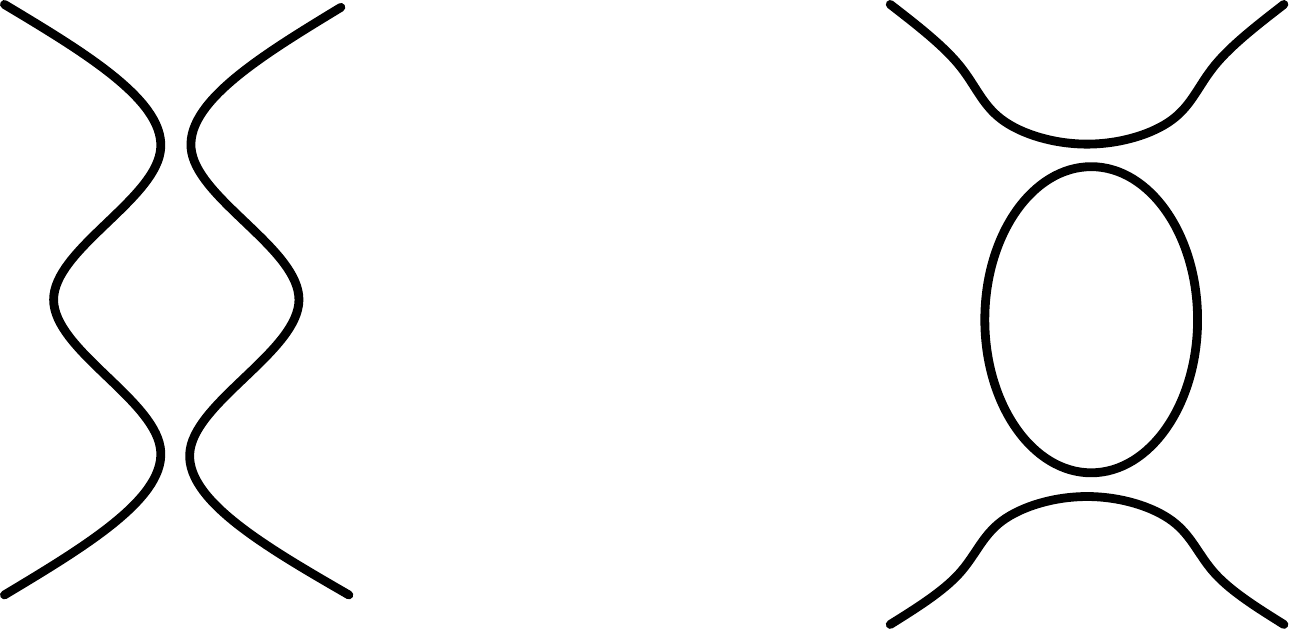}
\end{overpic}
\caption[Fig1-1]{Tensor structure of the divergent diagrams shown in Fig.~\ref{DivDiagrams}.}
\label{TensorStructureUGr1}
\end{figure}
In Fig.~\ref{TensorStructureUGr1} we redraw the two diagrams of Fig.~\ref{DivDiagrams} in the new notation. Taking the sum over these two diagrams, we obtain the four-point Green's function to order $\vkappa^2$ (with amputated external legs):
\begin{equation}
\big\langle\,U_a\,\thickbar{U}_b\,V_c\,\thickbar{V}_d\,\big\rangle=2\pi\vkappa\,\delta_{ab}\delta_{cd}
+2\pi\vkappa^2n\left(\delta_{ab}\delta_{cd}-\frac{1}{n}\delta_{ac}\delta_{bd}\right)
\ln\bigg(\frac{|p|}{\Lambda}\bigg)+\mathcal{O}(\vkappa^3).
\label{FourPointGreensFunctionUGr1}
\end{equation}
The loop contribution is proportional to the transverse  projector. The meaning of this is as follows. Strictly speaking, the ungauged Lagrangian~(\ref{LagrangianGNgaugedCP}), as it stands, is not stable under renormalization. The reason is that one could add to it  another term $2\pi\widetilde{\vkappa} |U_aV_a|^2$ with a new coupling constant $\widetilde\vkappa$ without breaking any symmetries. The formula (\ref{FourPointGreensFunctionUGr1}) then implies that only the transverse combination of the two terms gets renormalized.

As a result, the one-loop $\beta$-function is
\begin{empheq}[box = \fcolorbox{black}{white}]{align}
\label{BetaFunctionCP}
\quad \beta^{\mathsf{CP}}_\vkappa=-n\,\vkappa^2+\mathcal{O}(\vkappa^3). \quad
\end{empheq}

\subsection{$\beta$-function in the $\mathsf{Gr}(m,n)$ case.} 
The above result  can be easily generalized to the case of arbitrary $m$. Now the fields are $n\times m$ and $m\times n$ matrices, so this time we need to introduce two types of indices. The corresponding Gross-Neveu Lagrangian is
\bea
\label{LagrangianGNgauged}
\mathcal{L}=2\,\mathrm{Tr}\Big(V\thickbar\partial U+\thickbar{U}\partial\thickbar{V}\Big)+\frac{\vkappa}{2\pi}\,\mathrm{Tr}\big(J_{\mathsf{sl}}\thickbar{J}_{\mathsf{sl}}\big)\,,\qquad\frac{1}{2\pi}J_{\mathsf{sl}}:=\mu_{\mathsf{sl}}=UV. 
\eea
The propagators are diagonal with respect to both types of indices and their coordinate dependence is the same as in the case $m=1$. The interaction can be written in the form
\begin{empheq}[box = \fcolorbox{black}{white}]{align}
\label{InteractionUGrm}
\quad
\mathcal{L}_{\mathrm{int}}=\frac{\vkappa}{2\pi}\,\mathrm{Tr}\big(J_{\mathsf{sl}}\thickbar{J}_{\mathsf{sl}}\big)=2\pi\vkappa \left(U_{a\alpha}V_{\alpha b}\thickbar{V}_{b\beta}\thickbar{U}_{\beta a}\right).\quad 
\end{empheq}
Here and hereafter the Greek indices run from $1$ to $m$.

Again, we need to calculate the four-point Green's function $\big\langle\,U_{a\alpha }\,\thickbar{U}_{\beta b }\,V_{\gamma c}\,\thickbar{V}_{d\zeta}\,\big\rangle$. The key point is that the tensor structure of the interaction with respect to Greek indices generalizes the graphic  notation of Fig.~\ref{VertexUGr1} in a trivial way. Indeed, one can use the graphic notation set up in the $\mathsf{Gr}(1,n)$ case with the extra assumption that Greek indices run over the red and blue lines  independently. In this framework all diagrams contain the prefactor $\delta_{\alpha\gamma}\delta_{\beta\zeta}$. The four-point function differs from~(\ref{FourPointGreensFunctionUGr1})  just by this trivial prefactor, so that for arbitrary $m$ one gets  the same answer for the one-loop $\beta$-function:
\begin{empheq}[box = \fcolorbox{black}{white}]{align}
\label{BetaFunctionUGr1}
\quad \beta^{\mathsf{Gr}}_\vkappa=-n\,\vkappa^2+\mathcal{O}(\vkappa^3). \quad
\end{empheq}

\subsection{$\beta$-function in the $\mathsf{OGr}(m,n)$ case.} 
\begin{figure}[h!]
\centering
\begin{overpic}[scale=0.5,unit=0.8mm]{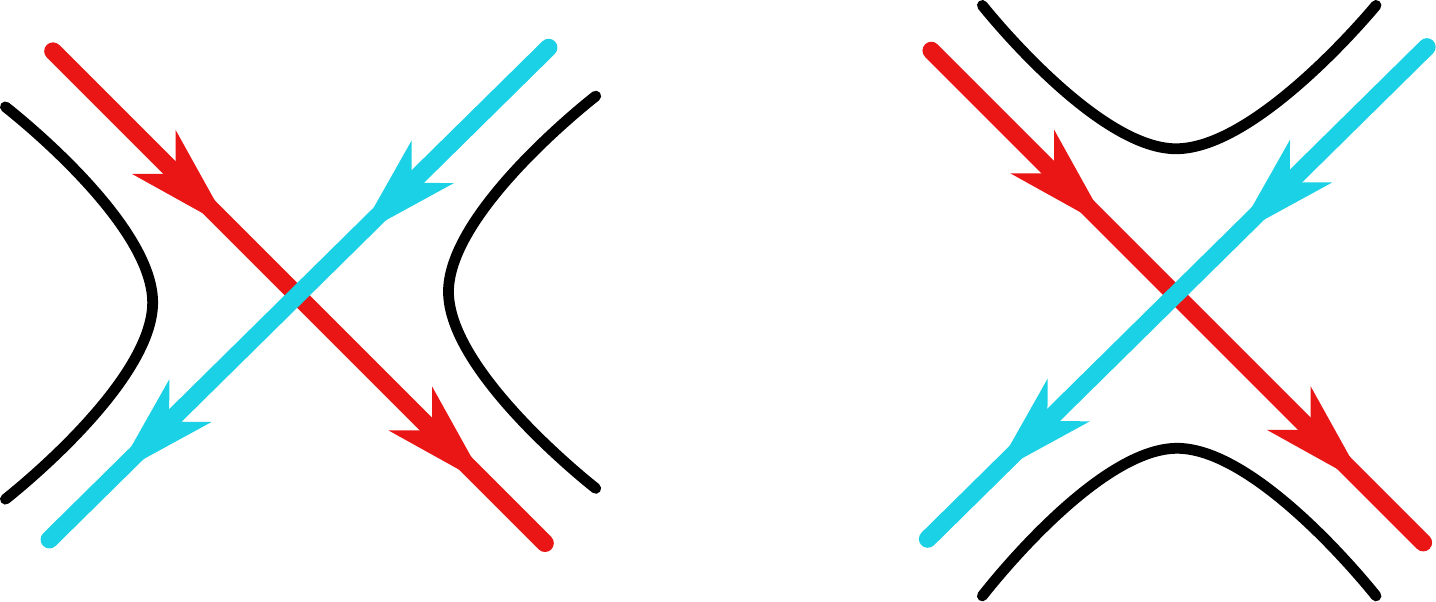}
\put(-8,35){\raisebox{.5pt}{\textcircled{\raisebox{-.9pt}{\large+}}}}
\put(50,35){\raisebox{.5pt}{\textcircled{\raisebox{-.9pt}{$\large-$}}}}
\end{overpic}
\caption[Fig1-1]{Two types of vertices.}
\label{VertOGr}
\end{figure}

It is easy to generalize the calculations of the previous section to the case of orthogonal and symplectic Grassmannians. Let us start with the orthogonal case.  
The Lagrangian has a structure similar to~(\ref{LagrangianGNgauged}),
\bea
\label{SymplecticLagrangianGNgauged}
\mathcal{L}=2\,\mathrm{Tr}\Big(V\mathds{h}_n\thickbar\partial U+\thickbar{U}\mathds{h}_n\partial\thickbar{V}\Big)+{\vkappa\over 4
\pi}\,\mathrm{Tr}\big(J_{\mathsf{o}}\thickbar{J}_{\mathsf{o}}\big),\qquad\frac{1}{2\pi}J_{\mathsf{o}}:=\mu_{\mathsf{o}},
\eea
and the moment map is given by formula~(\ref{SOmommap2}). The propagators are
\bea
\label{PropagatprsOGr1}
\Big\langle U_{a\alpha}(z,\thickbar{z}\,)V_{\beta b}(w,\thickbar{w}\,)\Big\rangle_0 = \frac{\delta_{\alpha\beta}\,\mathds{h}_{ab}}{2\pi(z-w)},\quad \Big\langle \thickbar{U}_{\alpha a}(z,\thickbar{z}\,)\thickbar{V}_{b\beta}(w,\thickbar{w}\,)\Big\rangle_0 = \frac{\delta_{\alpha\beta}\,\mathds{h}_{ab}}{2\pi(\thickbar{w}-\thickbar{z}\,)},
\eea
where $\mathds{h}_{ab}$ denotes the matrix elements of $\mathds{h}_{n}$ (recall that $\mathds{h}_{n}$ is symmetric and $\mathds{h}_{n}^2=\mathds{1}_n$). 
In terms of field components the interaction can be written as
\begin{empheq}[box = \fcolorbox{black}{white}]{align}
\label{InteractionOGrm}
\quad
\mathcal{L}_{\mathrm{int}}={\vkappa\over 4\pi}\,\mathrm{Tr}\big(J_{\mathsf{o}}\thickbar{J}_{\mathsf{o}}\big)=2\pi\vkappa\, \Big(U_{a\alpha}V_{\alpha b}\thickbar{V}_{b\beta}\thickbar{U}_{\beta a}\,-\,U_{a\alpha}V_{\alpha b}\thickbar{V}_{a\beta}\thickbar{U}_{\beta b}\Big).\quad 
\end{empheq}
By the same argument as in the previous section the $\beta$-function does not depend on~$m$. This means that one can set $m=1$ and use the Lagrangian with Greek indices suppressed. 
The whole difference with the case of a unitary Grassmannian lies in the presence of an additional vertex. Let us introduce similar graphic notations, see Fig.~\ref{VertOGr}. 
Here the two vertices correspond to the two terms in~(\ref{InteractionOGrm}) (the sign shown in the circle corresponds to the sign in front of the corresponding term in the Lagrangian).
\begin{figure}[h!]
\centering
\begin{overpic}[scale=0.7,unit=0.8mm]{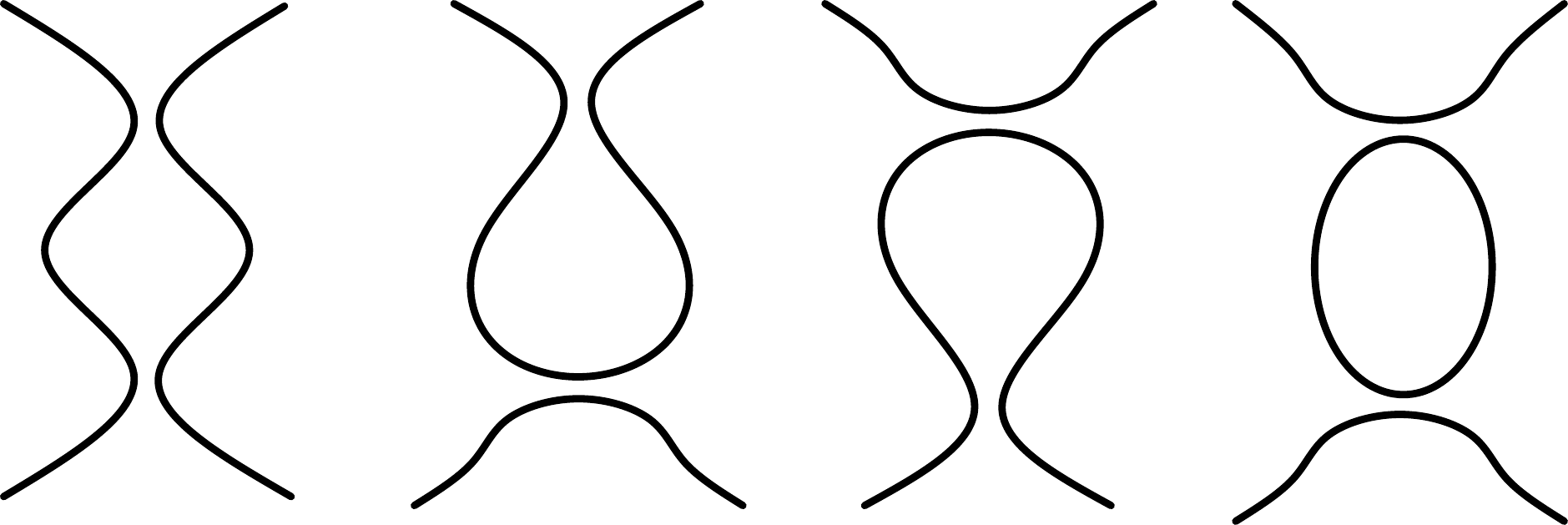}
\end{overpic}
\caption[Fig1-1]{Tensor structure of the divergent diagrams.}
\label{TensorStructureOGr1}
\end{figure}
One also has two types of divergent diagrams (as in Fig.~\ref{DivDiagrams}) each of which is divided into four diagrams according to the flow of flavor lines (these are shown in Fig.~\ref{TensorStructureOGr1}). 
Summing over these eight diagrams, we get
\begin{equation}
\big\langle\,\thickbar{U}_a\,U_b\,\thickbar{V}_c\,V_d\,\big\rangle=2\pi(\delta_{ab}\delta_{cd}-\delta_{ad}\delta_{cb})\Bigg(\vkappa
+
\vkappa^2(n-2)\ln\bigg(\frac{|p|}{\Lambda}\bigg)\Bigg)+\mathcal{O}(\vkappa^3).
\label{FourPointGreensFunctionOGr1}
\end{equation}
Note that all diagrams proportional to $\mathds{h}_{ac}\mathds{h}_{bd}$ cancel each other. The reason for this is the $\mathbb{Z}_2\times\mathbb{Z}_2$ symmetry of the action, which maps $U\rightarrow -V,~V\rightarrow U$ and $\thickbar{U}\rightarrow-\thickbar{V},~\thickbar{V}\rightarrow\thickbar{U}$.
It implies that the four-point Green's function (\ref{FourPointGreensFunctionOGr1}) is skew-symmetric with respect to $a\leftrightarrow c$ and $b\leftrightarrow d$. 
 From~(\ref{FourPointGreensFunctionOGr1}) one can read off the one-loop $\beta$-function of the $\mathsf{OGr}(m,n)$ model:
\begin{empheq}[box = \fcolorbox{black}{white}]{align}
\label{BetaFunctionOGr1}
\quad \beta^{\mathsf{OGr}}_\vkappa=-(n-2)\,\vkappa^2+\mathcal{O}(\vkappa^3).\quad 
\end{empheq}

\subsection{$\beta$-function in the $\mathsf{SGr}(m, 2n)$ case.} 
Here we consider the Gross-Neveu Lagrangian 
\bea
\label{SymplecticLagrangianGNgauged}
\mathcal{L}=2\,\mathrm{Tr}\Big(V\omega_{2n}\thickbar\partial U+\thickbar{U}\omega_{2n}\partial\thickbar{V}\Big)+{\vkappa\over 4
\pi}\,\mathrm{Tr}\big(J_{\mathsf{sp}}\thickbar{J}_{\mathsf{sp}}\big),\qquad\frac{1}{2\pi}J_{\mathsf{sp}}:=\mu_{\mathsf{sp}},
\eea
with the moment map (\ref{MomentMapSp}). The basic propagators differ from the ones of the previous section by a replacement $\mathds{h}_{ab}\rightarrow(-\omega_{ab})$, where $\omega_{ab}$ denotes the matrix elements of the $\omega_{2n}$, i.e.
\bea
\label{PropagatprsSGr1}
\Big\langle U_{a\alpha}(z,\thickbar{z}\,)V_{\beta b}(w,\thickbar{w}\,)\Big\rangle_0 = \frac{-\delta_{\alpha\beta}\,\omega_{ab}}{2\pi(z-w)},\quad \Big\langle \thickbar{U}_{\alpha a}(z,\thickbar{z}\,)\thickbar{V}_{b\beta}(w,\thickbar{w}\,)\Big\rangle_0 = \frac{-\delta_{\alpha\beta}\,\omega_{ab}}{2\pi(\thickbar{w}-\thickbar{z}\,)}.
\eea
Here we use that $\omega_{2n}^2=-\mathds{1}_{2n}$. Recall that $\omega_{2n}$ is skew-symmetric. One can check by a direct calculation that in this case the interaction  differs from   (\ref{InteractionOGrm}) by the sign in front of the first vertex, i.e.
\begin{empheq}[box = \fcolorbox{black}{white}]{align}
\label{InteractionSGrm}
\quad
\mathcal{L}_{\mathrm{int}}=-2\pi\vkappa\, \Big(U_{a\alpha}V_{\alpha b}\thickbar{V}_{b\beta}\thickbar{U}_{\beta a}\,+\,U_{a\alpha}V_{\alpha b}\thickbar{V}_{a\beta}\thickbar{U}_{\beta b}\Big).\quad 
\end{empheq}
As a result, the four-point function is proportional to $-(\delta_{ab}\delta_{cd}+\delta_{ad}\delta_{cb}).$ As in the previous case, one can assume that $m=1$ in calculations and there is a $\mathbb{Z}_2\times\mathbb{Z}_2$ symmetry of the theory mapping $U\leftrightarrow V$ and $\thickbar{U}\leftrightarrow\thickbar{V}$, which leads to the cancellation of all diagrams proportional to $\omega_{ac}\omega_{bd}$. 

One can see that the one-loop $\beta$-function in the $\mathsf{SGr}(m, 2n)$ case is obtained by a switch of sign in front of the second and third diagrams in Fig.~\ref{TensorStructureOGr1}. One also needs to add an extra sign in each diagram due to the replacement $\mathds{h}_{2n}\rightarrow(-\omega_{2n})$ in the propagators. As a result, we get
\begin{empheq}[box = \fcolorbox{black}{white}]{align}
\label{BetaFunctionSGr}
\quad \beta^{\mathsf{SGr}}_\vkappa=-2(n+1)\,\vkappa^2+\mathcal{O}(\vkappa^3). \quad
\end{empheq}

\subsection{Relation to Chern classes.}

In this section we interpret the results of the previous sections from a geometric standpoint, using the well-known relation of sigma model beta function to the Ricci tensor of the target space. In doing so, we  specialize to the K\"ahler case, i.e. to the case of Hermitian symmetric spaces shown in Table~\ref{hermsymmspace}. As can be seen from the explicit expressions~(\ref{orthomnlagr}) and~(\ref{lagrsympl2}), in our models the $B$-field is proportional to the fundamental Hermitian form of the metric, so that in K\"ahler cases it is closed, i.e. $H=dB=0$. In this class of models the general one-loop $\beta$-function of the metric is given by the formula \cite{Friedan,ZJ}
\begin{equation}\label{betafuncgen}
\beta_{ij}(g)=
R_{ij}+(\text{higher loops}),
\end{equation}
where $g_{ij}$ is the sigma model metric and $R_{ij}$ the Ricci tensor. Additionally, we have focused on metrics of the form $g={\vkappa^{-1}} \hat{g}$ (here $\hat{g}$ is some reference metric), finding that only the parameter $\vkappa$ undergoes renormalization:
\bea
\beta_{ij}(g)=\dot{g}_{ij}=\dot{\left({1\over \vkappa}\right)} \hat{g}_{ij}=-{\dot{\vkappa}\over \vkappa^2}\,\hat{g}_{ij}:=-{\beta_{\vkappa}\over \vkappa^2}\,\hat{g}_{ij}\,.
\eea
Compatibility with~(\ref{betafuncgen}) then asserts that $\hat{g}$ is K\"ahler-Einstein:
\bea\label{KE}
R_{ij}=-{\beta_{\vkappa}\over \vkappa^2}\,\hat{g}_{ij}\,.
\eea

On top of that, by a well-known result, the de Rham cohomology class of the Ricci form is  proportional to the first Chern class of the target space
\cite{moroianu}, i.e.
\begin{equation}\label{c1}
[c_1]_{\mathrm{dR}}=\frac{1}{2\pi}[\mathrm{Ric}]_{\mathrm{dR}}. 
\end{equation}
Thus, knowing the cohomology class of the K\"ahler form of the metric $\hat{g}$ and the first Chern class $c_1$ allows one to compute $\beta_{\vkappa}$ from~(\ref{KE}). 
This fact can be used to cross-check  our $\beta$-function computations in the K\"ahler cases.

To be able to apply the above formulas~(\ref{KE})-(\ref{c1}), it is useful to recall how the metrics of our sigma models arise. In all cases, they are related to the metrics on the corresponding Grassmannians $\mathsf{Gr}(m, n)$: in the unitary case these are the metrics themselves, whereas in the orthogonal and symplectic cases the metrics are obtained by restriction. The metric on $\mathsf{Gr}(m, n)$, which we restrict,
is unique up to normalization, which is most conveniently described as follows. Let $\mathcal{S}^\vee$ be the dual of the tautological bundle over $\mathsf{Gr}(m, n)$, then the K\"ahler class may be written as $[\frac{1}{2\pi}\hat{g}]=\upalpha\, c_1(\mathcal{S}^\vee)$, 
where $\upalpha>0$. Geometrically, $\upalpha$ is the integral of the K\"ahler form over the elementary $\CP^1\subset \mathsf{Gr}(m, n)$. We may then write
\bea\label{Chernclassbeta}
c_1(\textrm{target space})=-{\beta_{\vkappa}\over \vkappa^2}\cdot\upalpha\cdot  c_1(\mathcal{S}^\vee)\,.
\eea

This formula is especially useful, since in practice the first Chern classes of all the  Grassmannians shown in Table~\ref{hermsymmspace} can be easily expressed in terms of $c_1(\mathcal{S}^\vee)$. We will now see this in explicit computations. For details of the setup used  see~\cite{eisenbud_harris_2016}, for example.

\subsubsection{Unitary Grassmannians.}
To compute the first Chern class of $\mathsf{Gr}(m,n)$ it is useful to express the tangent bundle in terms of the tautological one. The decomposition has the form:
\begin{equation}
\label{UniversalBundleDecomposition}
\mathrm{T}\mathsf{Gr}(m,n)\simeq\mathcal{S}^\vee\otimes\mathcal{Q}.
\end{equation}
Here $\mathcal{S}$ is the tautological bundle over the Grassmannian, $\mathcal{S}^\vee$ is its dual and $\mathcal{Q}\simeq\mathsf{C}^n\big/\mathcal{S}$ is the quotient bundle, so that  $c_1\left(\mathcal{S}^\vee\right)=c_1(\mathcal{Q})$. Thus,
\begin{equation}\label{c1Grmn}
c_1\left(\mathsf{Gr}(m,n)\right)=(n-m)\,c_1\left(\mathcal{S}^\vee\right)+m\,c_1(\mathcal{Q})=n\,c_1(\mathcal{S}^\vee)=h^\vee_{\mathsf{sl}(n)}\,c_1(\mathcal{S}^\vee).
\end{equation}
This may now be compared with~(\ref{Chernclassbeta}) to relate the $\beta$-function with the normalization $\upalpha$ of the metric. Indeed, it follows from~(\ref{BetaFunctionUGr1}) that $-{\beta_{\vkappa}\over \vkappa^2}=n$, so that~(\ref{Chernclassbeta}) together with~(\ref{c1Grmn}) imply $\upalpha=1$. On the other hand, we know from section~\ref{unitarygrass} that in this case the relevant Lagrangian is~(\ref{UGrassLagr}), so that the value $\upalpha=1$ corresponds to the normalization of the metric featuring in that Lagrangian. Hence, we can now compare the normalizations of all other metrics with the one in~(\ref{UGrassLagr}) to recover the values of $\upalpha$ in the orthogonal and symplectic cases.

\subsubsection{Grassmannian of real 2-planes.} 
Before passing to the general case, we consider the simplest example of an orthogonal Grassmannian:  $\mathsf{OGr}(1,n)\simeq Q$, which is a non-singular quadric hypersurface $Q$ in $M=\mathsf{CP}^{n-1}=\mathsf{Gr}(1,n)$. The following decomposition then holds (the adjunction formula \cite{GriffithsHarris}):
\begin{equation}
\label{Adjunction}
\mathrm{T}M\big|_Q\simeq\mathrm{T}Q\oplus\mathcal{N}_Q\,,
\end{equation}
where $\mathcal{N}_Q$ is the normal bundle to $Q\mysub \CP^{n-1}$. The quadric $Q$ is the set of zeroes of a section of the line bundle $\mathcal{O}(2)$ over $\CP^{n-1}$, so that the normal bundle is $\mathcal{N}_Q\simeq\mathcal{O}(2)$ restricted to $Q$. Besides, if $\mathcal{S}$ is the tautological bundle over $\CP^{n-1}$,  $\mathcal{S}^\vee \simeq \mathcal{O}(1)$.  
Calculating the first Chern class of both sides of (\ref{Adjunction}) we then get
\bea\label{c1Q}
c_1\left(\mathsf{OGr}(1,n)\right)=(n-2)\,c_1(\mathcal{S}^\vee)=h^\vee_{\mathsf{o}(n)}\,c_1(\mathcal{S}^\vee),
\eea
where $h^\vee$ is the dual Coxeter number from Table~\ref{CoxeterTable}. Again, this formula may be used together with~(\ref{Chernclassbeta}) to deduce the $\beta$-function. By comparing~(\ref{OGr1nLagr}) with the reference Lagrangian~(\ref{UGrassLagr}), one sees that in this case $\upalpha=1$. The beta function is then found from~(\ref{Chernclassbeta}) as $\beta_{\vkappa}=-(n-2)\,\vkappa^2$, in full accord with~(\ref{BetaFunctionOGr1}).

\subsubsection{Orthogonal and symplectic Grassmannians.}

Just as in the example above, the general idea is that one can consider the orthogonal Grassmanians $\mathsf{OGr}(m,n)$ and symplectic Grassmannians $\mathsf{SGr}(m,n)$ as submanifolds in ordinary Grassmannians $\mathsf{Gr}(m,n)$. The decomposition (\ref{Adjunction}) may then be generalized accordingly:
\begin{align}
\label{AdjunctionOGr}\mathrm{T}\mathsf{Gr}(m,n)\big|_{\mathsf{OGr}(m,n)}\simeq\mathrm{T}\mathsf{OGr}(m,n)\oplus\mathcal{N}_{\mathsf{OGr}(m,n)}
\end{align}
and analogously for symplectic Grassmannians. 
The embeddings $\mathsf{OGr}(m,n)\subset\mathsf{Gr}(m,n)$ and $\mathsf{SGr}(m,n)\subset\mathsf{Gr}(m,n)$ can be realized as zeroes of sections of the vector bundles $\mathrm{Sym}^2\mathcal{S}^\vee$ and $\bigwedge^2\mathcal{S}^\vee$ respectively.
Thus, 
\begin{equation}
\mathcal{N}_{\mathsf{OGr}(m,n)}\simeq\left(\mathrm{Sym}^2\mathcal{S}^\vee\right)\big|_{\mathsf{OGr}(m,n)}\qquad\text{and}\qquad \mathcal{N}_{\mathsf{SGr}(m,n)}\simeq\left(\bigwedge^2\mathcal{S}^\vee\right)\bigg|_{\mathsf{SGr}(m,n)}\,.
\end{equation}
By the splitting principle one can express the Chern classes of $\mathrm{Sym}^2\mathcal{S}^\vee$ and $\bigwedge^2\mathcal{S}^\vee$ in terms of the Chern classes of $\mathcal{S}^\vee$. The result of the calculation is that
\begin{equation}
c_1\left(\mathrm{Sym}^2\mathcal{S}^\vee\right)=(m+1)c_1\left(\mathcal{S}^\vee\right)\qquad\text{and}\qquad c_1\left(\bigwedge^2\mathcal{S}^\vee\right)=(m-1)c_1\left(\mathcal{S}^\vee\right).
\end{equation}
Computing the first Chern class of (\ref{AdjunctionOGr}) and of the analogous decomposition for symplectic Grassmannians, we get
\bear
&&c_1\left(\mathsf{OGr}(m,n)\right)=(n-m-1) \,c_1\left(\mathcal{S}^\vee\right)\,,
\\ &&c_1\left(\mathsf{SGr}(m,n)\right)=(n-m+1)\,c_1\left(\mathcal{S}^\vee\right)\,.
\eear
In the case of $\mathsf{OGr}(1,n)$ the result  coincides with~(\ref{c1Q}). In the maximal cases -- the variety of projective pure spinors $\mathsf{OGr}(m,2m)$ (cf.~\cite{BerkovitsNekrasov}) and the Lagrangian Grassmannian $\mathsf{SGr}(m,2m)$ -- the corresponding Chern classes are 
\bear\label{OGrm2mc1}
&&c_1\left(\mathsf{OGr}(m,2m)\right)=(m-1)\,c_1\left(\mathcal{S}^\vee\right)={h^\vee_{\mathsf{o}(2m)} \over 2} \, c_1\left(\mathcal{S}^\vee\right)\,,\quad\quad \\ \label{SGrm2mc1}&&c_1\left(\mathsf{SGr}(m,2m)\right)=(m+1)\,c_1\left(\mathcal{S}^\vee\right)=h_{\mathsf{sp}(2m)}^\vee\,c_1\left(\mathcal{S}^\vee\right)\,,
\eear
where $h^\vee$ are the dual Coxeter numbers of the respective symmetry algebras, as summarized in Table~\ref{CoxeterTable}.

As it follows from the discussions in sections~\ref{maxOGrass} and~\ref{maxSGrass}, for both $\mathsf{OGr}(m,2m)$ and $\mathsf{SGr}(m,2m)$ the normalization of the sigma model metric is $1\over 2$ times that of the unitary Grassmannian~(\ref{UGrassLagr}), so that here $\upalpha={1\over 2}$. The general formula~(\ref{Chernclassbeta}), together with~(\ref{OGrm2mc1}) and  (\ref{SGrm2mc1}), then implies that in both cases $\beta_{\vkappa}$ matches the  values~(\ref{BetaFunctionOGr1}) and~(\ref{BetaFunctionSGr})  calculated above.

\section{Conclusion and outlook}

In the present paper we formulated sigma models of orthogonal and symplectic Grassmannians as generalized Gross-Neveu models. This is a natural generalization of earlier work on unitary Grassmannians. 
As a first application of the new formalism, we calculated the one-loop beta functions, showing that they are proportional to the dual Coxeter numbers of the respective symmetry algebras.

An important relation that is visible through the results of the present paper but was not  elaborated above is the relation between sigma models admitting a GN formulation and nilpotent orbits of complex Lie groups. Indeed, recall that in all of the cases considered the relevant moment map $\mu$ is nilpotent, cf.~(\ref{munilpeq}), (\ref{munilpeq2}), (\ref{munilpeq3}). This is not a coincidence, and the relation was discussed to a certain extent in~\cite{BykovNilp} in the $\tSU(n)$ case. Generally speaking, the Jordan type of the orbit characterizes the target space of the resulting sigma model. Moreover,  nilpotent orbits are known to admit quiver formulations~\cite{Nakajima, KobakSwann}, which apparently is the ultimate reason  why the GN-type formulation exists.

We have shown that the metrics of the resulting sigma models are K\"ahler if and only if the target space is a Hermitian symmetric space. It is well-known that K\"ahler models admit $\mathcal{N}=(2,2)$ SUSY completions, and it would be interesting to construct them in the GN formalism, extending the results for the unitary Grassmannians~\cite{BykovSUSY}. In general, in quantum theory one is inevitably led to consider fermionic extensions of the models (either supersymmetric or not), since fermions are needed for the cancellation of chiral anomalies, cf. the discussion in~\cite{BykovRiemann}.

Another direction where our methods could be extended include flag manifold target spaces. A generalization to the  case of exceptional symmetry groups is of significant interest as well. Last but not least, in our exposition above we concentrated on the formulation of the models in terms of the GN variables, but  paid virtually no attention to the integrable structure of the resulting models. We plan to return to these questions in the future.

\vspace{1cm}
\textbf{Acknowledgments.} Sections 1-2 were written with the support of the Foundation for the Advancement of Theoretical Physics and Mathematics ``BASIS''. Sections 3-7 were supported by the Russian Science Foundation grant № 22-72-10122 (\href{https://rscf.ru/en/project/22-72-10122/}{\emph{https://rscf.ru/en/project/22-72-10122/}}). We would like to thank M.~Kontsevich, G.~Korchemsky,  M.~Markov, N.~Mekareeya, A.~Rosly and A.~Smilga for discussions and E.~Sharpe for comments on the manuscript.

\vspace{1cm}
 
\appendix

\section{$\mathsf{O}$- and $\mathsf{Sp}$-Grassmannians as homogeneous spaces}\label{homspaceapp}

For most of the paper we used the complex definition of  Grassmannians as sets of isotropic planes in a given ambient space $\CC^n$. As mentioned in the introduction, they are also homogeneous spaces~(\ref{UGr}),~(\ref{OGr}),~(\ref{SGr}) of the respective compact symmetry groups. In the present Appendix we recall the relation between these two definitions.

\subsection{$\mathsf{OGr}(m, n)$ as homogeneous spaces.}\label{OGrHomSpacesec}
We start with the orthogonal Grassmannians. Here one has the identification
\bea\label{orthohomspace}
\mathsf{OGr}(m, n)=\frac{\tO(n)}{\tU(m)\times \tO(n-2m)}\,,
\eea
where $\tU(m)$ is embedded into the diagonal $\tO(2m)\subset \tO(n)$. In particular, the Grassmannian of real 2-planes  arises for $m=1$, whereas the Grassmannian of orthogonal complex structures in $\mathsf{R}^{2m}$ corresponds to $n=2m$. Both of these are  symmetric spaces, see Table~\ref{hermsymmspace}.

There are two steps in proving the above representation~(\ref{orthohomspace}). First, we should show that the compact subgroup $\tO(n)=\tO(n, \CC) \bigcap \tU(n)$ acts transitively on the isotropic planes. In proving this, we will assume that the symmetric form $\mathds{h}_n$ is chosen as in~(\ref{symmformsplit}). Take an arbitrary isotropic plane parametrized by an $m\times n$ matrix $U$ and choose an orthonormal basis in it w.r.t. the standard scalar product in $\CC^n$, so that $\thickbar{U}U=\mathds{1}_m$. One can embed this matrix $U$ in an $n\times n$ unitary orthogonal matrix
\bea\label{Uembed}
g=\left(U\; \mathds{h}_n U^\ast\;Y\right)\,,
\eea
where $Y$ is the matrix of $n-2m$ complimentary orthonormal vectors, satisfying
\bear\label{Yortho}
&&{\thickbar{Y}U=0},\quad\quad Y^t \mathds{h}_n U=0\\ \label{Ynorm} &&\textrm{and} \quad\quad \thickbar{Y}Y=\mathds{1}_{n-2m}\,.
\eear
Indeed, pick an arbitrary matrix $Y_0$ satisfying ${\thickbar{Y}_0U=0}$ and of maximal rank $n-2m$. If $Y_0^t \mathds{h}_n U=X$ ($\neq 0$) we may set $Y=Y_0-\mathds{h}_n U^\ast X^t$, which would satisfy~(\ref{Yortho}). Orthonormalizing the vectors in $Y$, we ensure that~(\ref{Ynorm}) is satisfied as well.

As a result, $g$ is unitary, $\thickbar{g}g=\mathds{1}_n$, and also orthogonal:
\bea
g^t\mathds{h}_n g=\mathds{h}_n\,\quad\quad \Longrightarrow\quad\quad g\in \tO(n)\,.
\eea
If one now has another isotropic plane parametrized by the matrix $U'$, we can equally embed it in a matrix $g'\in \tO(n)$. The two matrices are thus related by a group element
\bea
g'=g_0 g,\quad\quad g_0\in \tO(n) \quad\quad \Longrightarrow\quad\quad U'=g_0 U\,,
\eea
so that indeed $\tO(n)$ acts transitively on such planes.

The second step is to find the stabilizer of $m$ vectors $u_1, \cdots, u_m$ in $\CC^n$ satisfying the isotropy constraint~(\ref{orthocons}). As before, we will take the symmetric form~(\ref{symmformsplit}), and the $m$ vectors spanning the isotropic subspace as\footnote{Here $E_j$ are the standard unit vectors with components $(E_j)_k=\delta_{jk}$.}  $u_j=E_j$, $j=1, \ldots, m$. We are looking for matrices $\mathbf{g}\in \tO(n, \CC)$ stabilizing this subspace and, in addition, belonging to the compact subgroup $\tO(n)=\tO(n, \CC) \bigcap \tU(n)$. One easily finds that these are matrices of the form
\bea
\mathbf{g}=\begin{pmatrix}
    \mathbf{g}_1 & 0 & 0\\
    0& \left(\mathbf{g}_1^t\right)^{-1} &0\\
    0 & 0 &  \mathbf{g}_2
\end{pmatrix}\,\in \tO(n)\,,
\eea
where $\mathbf{g}_1\in \tU(m)$ and $\mathbf{g}_2\in \tO(n-2m)$. This therefore proves~(\ref{orthohomspace}).

\subsubsection{$\tSO$ versus $\tO$. }

As a slight digression, let us study when the connected subgroup $\tSO(n)\mysub \tO(n)$ acts transitively on the orthogonal Grassmannian. First, take $n=2$ and $m=1$, where the isotropy constraint can be solved by  taking either $u=\begin{pmatrix}
1\\
0
\end{pmatrix}$ or $\widetilde{u}=\begin{pmatrix}
0\\
1
\end{pmatrix}$. The two are related by the matrix $g_0=\begin{pmatrix}
0 & 1\\
1 & 0
\end{pmatrix}$ with $\mathrm{Det}(g_0)=-1$, so that $g_0\in \tO(2)$. Thus here $\mathsf{OGr}(1,2)={\tO(2)\over \tU(1)}=\mathbb{Z}_2$ is a set of two points.

For higher $m$, one could take the vectors $u_j=E_j$, $j=1, \ldots, m$, or $\widetilde{u}_j:=E_{m+j}$, $j=1, \ldots, m$, or a mixture of those. Notice, however, that whenever $n\neq 2m$, one can rotate $\widetilde{u}_j$ into $u_j$ by a transformation from $\tSO(n)$. For $n=3$ one simply takes the matrix $g_1=\begin{pmatrix}
g_0 & 0\\
0  & -1
\end{pmatrix}$ with $\mathrm{Det} (g_1)=1$, and for higher $n$ its suitable embedding. Thus, in these cases $\tSO(n)$ acts transitively on the set of isotropic planes.

The only remaining case is $n=2m$. Here one can rotate pairs of vectors $\widetilde{u}_1, \widetilde{u}_2$ into $u_1, u_2$ by choosing the permutation matrix $g_2=\begin{pmatrix}
0 & \mathds{1}_2\\
\mathds{1}_2  & 0
\end{pmatrix}$ with $\mathrm{Det} (g_2)=1$, or its appropriate embedding. Therefore one should differentiate between two cases: when the number of $\widetilde{u}_j$'s is even or odd.  One can transfer from one case to the other only by an $\tO(n)$, but not an $\tSO(n)$ transformation,  and, as a result, the orthogonal Grassmannian has two connected components:
\bea
\mathsf{OGr}(m, 2m)=\mathsf{OGr}^{+}(m, 2m)\bigsqcup \mathsf{OGr}^{-}(m, 2m)\,,
\eea
where each one is $\mathsf{OGr}^{\pm}(m, 2m)\simeq {\tSO(2m)\over \tU(m)}$.

\subsection{$\mathsf{SGr}(m, 2n)$ as a homogeneous space.} \label{SGrHomSpacesec}

We pass over to the description of symplectic Grassmannians as homogeneous spaces of the compact group $\tSp(2n)$. More precisely, 
one has the identification
\bea\label{symplhomspace}
\mathsf{SGr}(m, 2n)=\frac{\tSp(2n)}{\tU(m)\times \tSp(2(n-m))}\,.
\eea
Here $\tU(m)$ is embedded into the diagonal $\tSp(2m)\subset \tSp(2n)$. As an elementary check, one can compute dimensions to arrive at the correct value, see~(\ref{symplgrdim}).

First, we prove that the compact group $\tSp(2n)$ acts transitively on the space of isotropic planes. As usual, we start with the matrix $U$ comprising the $m$ linearly independent vectors in the plane. We assume these vectors are orthonormal, so that $U^t \omega_{2n} U=0$ and $\thickbar{U}U=\mathds{1}_m$. Given $U$, we construct a matrix $g\in \tSp(2n)$ as follows:
\bea\label{Uembed2}
g=(U\;\omega_{2n}^{-1} U^\ast\; Y)\,,
\eea
where $Y$ is a complimentary set of vectors, such that $Y^t\omega_{2n} U=0$ and $\thickbar{Y}U=0$. By construction, $\thickbar{g}g=\mathds{1}_{2n}$, so $g$ is unitary, and $g^t\omega_{2n} g=\omega_{2n}$, so that $g$ is symplectic. In other words, $g\in\tSp(2n)$.

If one has another isotropic $m$-plane parametrized by the matrix $U'$, in exactly the same way we construct the group element $g'\in \tSp(2n)$. As a result, the two group elements are related by an $\tSp(2n)$-transformation:
\bea
g'=g_0 g,\quad\quad g_0\in \tSp(2n) \quad\quad \Longrightarrow\quad\quad U'=g_0 U\,.
\eea
Thus, transitivity is proven.

To calculate the stabilizer, it will be convenient to split the symplectic forms in two parts
\bea\label{symplcan1}
\omega_{2n}=\begin{pmatrix}
    \omega_{2m} & 0\\
    0 & \omega_{2n-2m}
\end{pmatrix}
\eea
as in~(\ref{omegasymplsplit}). We may now take $u_j=E_j$, $j=1, \ldots, m$ as the basis in the isotropic $m$-plane. Matrices $\mathbf{g}$ stabilizing this subspace and belonging to the compact subgroup $\tSp(2n)=\tSp(2n, \CC)\bigcap \tU(2n)$ are of the form
\bea
\mathbf{g}=\begin{pmatrix}
    \mathbf{g}_1 & 0 & 0\\
    0& (\mathbf{g}_1^t)^{-1} &0\\
    0 & 0 &  \mathbf{g}_2
\end{pmatrix}\,,
\eea
where $g_1\in \tU(m)$ and $g_2\in \tSp(2n-2m)$, which therefore proves~(\ref{symplhomspace}).

Let us again consider the interesting limiting cases. Clearly, $m=n$ corresponds to the Lagrangian Grassmannian. The case $m=1$ is more curious. Here the isotropy constraint~(\ref{symplcons})  is trivial, so that
\bea
\mathsf{SGr}(1, 2n)\simeq \CP^{2n-1}\,.
\eea
To deduce this from the quotient space representation~(\ref{symplhomspace}), note that for $4n-1$-dimensional spheres one has the representation
\bea\label{sphere4n}
\mathsf{S}^{4n-1}\simeq \frac{\tSp(2n)}{\tSp(2(n-1))}\,.
\eea
This is a quaternionic analogue of the representation $\mathsf{S}^{2n-1}\simeq{\tSU(n)\over \tSU(n-1)}$ for odd-dimensional spheres. In both cases, the quotients represent homogeneous but not symmetric spaces. According to (\ref{symplhomspace}), the symplectic Grassmannian~$\mathsf{SGr}(1, 2n)$ involves an additional quotient of the sphere~(\ref{sphere4n}) by $\tU(1)$, which is the Hopf fibration and leads to the projective space $\CP^{2n-1}$.

\section{Invariant metrics}\label{invmetrapp}

Here we shall construct the most general invariant metrics on the orthogonal and symplectic Grassmannians. As we will show, generically there is a family of metrics, and the metric that arises out of the Gross-Neveu formulation corresponds to a special point in that family.

\subsection{Orthogonal Grassmannians $\mathsf{OGr}(m, n)$.}\label{orthoalgapp} 

Here we return to the presentation~(\ref{orthohomspace}) of the orthogonal Grassmannian as a homogeneous space of $\tO(n)$. Accordingly we perform the Lie algebra decomposition
\bear\label{ondecomp}
&&\mathsf{o}(n)=\mathsf{h}\oplus \mathsf{m}\,,\quad\quad \textrm{where}\quad\quad \\ \nonumber
&&\mathsf{h}=\mathsf{u}(m)\oplus \mathsf{o}(n-2m)\,,\quad\quad \mathsf{m}=\mathsf{h}^{\perp}\,,
\eear
where by $\mathsf{h}^{\perp}$ we mean the orthogonal complement to $\mathsf{h}$  w.r.t. the Killing metric. Since $\mathsf{o}(n)$ is the space of skew-symmetric $n\times n$-matrices, the metric is simply the trace form on such matrices. Besides, $[\mathsf{h}, \mathsf{m}]\subset \mathsf{m}$, so that the subalgebra $\mathsf{h}$ is represented on $\mathsf{m}$. Moreover, over the real numbers $\mathsf{m}$ splits into two irreducible representations ($[\mathsf{h}, \mathsf{m}_i]\subset \mathsf{m}_i$):
\bea\label{mirrepdec}
\mathsf{m}=\mathsf{m}_1\oplus \mathsf{m}_2\,.
\eea
Notice that the extreme cases -- the symmetric spaces $\tSO(n)\over \tSO(2)\times \tSO(n-2)$ and~$\tSO(2m)\over \tU(m)$ -- correspond to $\mathsf{m}_2=0$ and $\mathsf{m}_1=0$ respectively. The above decomposition is more vividly shown in Fig.~\ref{ondecompfig}.

\begin{figure}
\begin{center}
 \begin{tikzpicture}[every node/.style={text height=3.3ex,text width=1.8em}]
\put(-115,-7) {$\mathsf{o(n)}=$};
\put(-60,20) {$\mathsf{u}(m)\oplus \mathsf{m_2}$};
\put(5,-32) {$\mathsf{o}(n-2m)$};
\put(10,20) {$A\in\mathsf{m_1}$};
\put(-60,-32) {$-A^t\in\mathsf{m_1}$};
\matrix[matrix of math nodes,
        left delimiter=(,
        right delimiter=),
        nodes in empty cells] (m)
{
       &  &   &    \\
  &         &   &             \\
       &   &   &             \\
       &   &   &             \\
};
\path[draw, line width = 0.3mm] (m-1-2.north east) -- (m-4-2.south east);
\path[draw, line width = 0.3mm] (m-2-1.south west) -- (m-2-4.south east);
    \end{tikzpicture}
    \end{center}
\caption{Decomposition~(\ref{ondecomp}) of the Lie algebra $\mathsf{o(n)}$. Any element of $\mathsf{m_1}$ is determined by $A\in\mathrm{Hom}(\CC^{n-2m}, \CC^{2m})$.}
\label{ondecompfig}
    \end{figure}

We wish to describe $\mathsf{m}_1$ and $\mathsf{m}_2$ more explicitly. To describe $\mathsf{m}_2$, pick a complex structure
\bea\label{compstruct}
\mathcal{J}=\mathds{1}_m\otimes i\sigma_2\,.
\eea
Then the stabilizer $\mathsf{u}(m)$ is defined as the set of transformations preserving it:
\bea
\mathsf{u}(m)=\Bigl\{\quad \alpha\in \mathsf{o}(2m),\quad\quad [\alpha, \mathcal{J}]=0\quad \Bigl\}\,.
\eea
It can be described explicitly as follows:
\bea\label{alphadef}
\alpha=\alpha_1\otimes \mathds{1}_2+\alpha_2\otimes i\sigma_2\,\quad\quad \textrm{where}\quad\quad \alpha_1^t=-\alpha_1,\quad \alpha_2^t=\alpha_2\,.
\eea
Accordingly, the orthogonal complement $\mathsf{m}_2$ is described by matrices\footnote{More invariantly, one arrives at the definition
\bea
\mathsf{m}_2=\Bigl\{\quad \beta\in \mathsf{o}(2m),\quad\quad \{\beta, \mathcal{J}\}=0\quad \Bigl\}\,.
\eea
This follows from~(\ref{compstruct}), (\ref{betadef}) by inspection, but the real reason is that $\mathsf{m}_2$ describes deformations of complex structures. That is, from the definition of complex structure, $\mathcal{J}^2=-\mathrm{Id}$, one finds $(\mathcal{J}+\beta)^2=\mathcal{J}^2+\{\beta, \mathcal{J}\}+\cdots=-1$, so that $\{\beta, \mathcal{J}\}=0$.}
\bea\label{betadef}
\beta=\beta_1\otimes \sigma_1+\beta_2\otimes \sigma_3\,\quad\quad \textrm{where}\quad\quad \beta_1^t=-\beta_1,\quad \beta_2^t=-\beta_2\,.
\eea

Next, we can define a complex structure on $\mathsf{m}_2$ by defining the holomorphic/anti-holomorphic subspaces of $(\mathsf{m}_2)_{\CC}$:
\bea\label{m2hol}
(\mathsf{m}_2)_{\pm}=\Bigl\{\quad \beta\otimes (\sigma_1\mp i \sigma_3)\,,\quad\quad \beta^t=-\beta\quad\Bigl\}\,.
\eea
It is easy to see that $(\mathsf{m}_2)_{\pm}$ are isotropic w.r.t. the trace form. We can also check that these subspaces are preserved by the action of the stabilizer $\mathsf{u}(m)$:
\bea
[\alpha_1\otimes \mathds{1}_2+\alpha_2\otimes i\sigma_2, \beta\otimes (\sigma_1\mp i \sigma_3)]=
\left([\alpha_1, \beta]-i \{\alpha_2, \beta\}\right)\otimes (\sigma_1\mp i \sigma_3)\,.
\eea
Thus, $(\mathsf{m}_2)_{\CC}=\mathsf{m}_{2+}+\mathsf{m}_{2-}$ splits into a sum of two (conjugate) irreps of $\mathsf{u}(m)$.

We will also pick a complex structure on $\mathsf{m}_1$. As shown in Fig.~\ref{ondecompfig}, $\mathsf{m_1}$ may be identified with the space of matrices $\mathrm{Hom}(\CC^{n-2m}, \CC^{2m})$ of size $(2m)\times(n-2m)$.  
The holomorphic/anti-holomorphic subspaces are then defined as follows:
\bea\label{m1hol}
\mathsf{m}_{1\pm}=\Biggl\{\quad v\otimes \begin{pmatrix}1\\ \pm i\end{pmatrix}\quad\Biggl\}\,,\quad\quad v\in \mathrm{Hom}(\CC^{n-2m}, \CC^{m})\,.
\eea
Again let us show that these subspaces are preserved by the action of $\mathsf{u}(m)$:
\bea
\left(\alpha_1\otimes \mathds{1}_2+\alpha_2\otimes (i\sigma_2)\right)\,v\otimes \begin{pmatrix}1\\ \pm i\end{pmatrix}= (\alpha_1\pm i \alpha_2)v\otimes \begin{pmatrix}1\\ \pm i\end{pmatrix}\,.
\eea
Thus, $\mathsf{u}(m)$ acts on the vector $v\in \mathrm{Hom}(\CC^{n-2m}, \CC^{m})$ by anti-Hermitian matrices, as it should.

Finally, we need the commutation properties\footnote{The commutation properties of $[\mathsf{m}_{2\pm}, \mathsf{m}_{2\pm}]$ and $[\mathsf{m}_{2\pm}, \mathsf{m}_{2\mp}]$ follow from the fact that $\tSO(2m)\over \tU(m)$ is a Hermitian symmetric space.} of $[\mathsf{m}_2, \mathsf{m}_1]$ and $[\mathsf{m}_1, \mathsf{m}_1]$. Consider the action of $\mathsf{m}_2$ on $\mathsf{m}_{1\pm}$:
\bea
(\beta_1 \otimes \sigma_1+\beta_2\otimes \sigma_3)\,v\otimes \begin{pmatrix}1\\ \pm i\end{pmatrix}= (\beta_2\pm i \beta_1)v \otimes \begin{pmatrix}1\\ \mp i\end{pmatrix}\,.
\eea
As for the last commutator, clearly, $[\mathsf{m}_1, \mathsf{m}_1]\subset \mathsf{o}(n-2m)\oplus (\mathsf{u}(m)\oplus \mathsf{m}_2)$, and we are interested in disentangling the $\mathsf{u}(m)$- and $\mathsf{m}_2$- components of the r.h.s. One gets (with hopefully self-evident notations)
\bear
&&[\mathsf{m}_{1\pm}, \mathsf{m}_{1\pm}]=(v_2v_1^t-v_1v_2^t)\otimes (\sigma_3\pm i \sigma_1)\,\subset \mathsf{m}_{2\pm}\,,\\
&&[\mathsf{m}_{1\pm}, \mathsf{m}_{1\mp}]=(v_2v_1^t-v_1v_2^t)\otimes \mathds{1}_2\mp (v_2v_1^t+v_1v_2^t)\otimes \sigma_2\,\subset \mathsf{u}(m)\,.
\eear

Collecting all of the above, we get the following relations:
\begin{empheq}[box = \fcolorbox{black}{white}]{align}
\nonumber
&[\mathsf{m}_{2\pm}, \mathsf{m}_{1\pm}]=0\,,\quad\quad [\mathsf{m}_{1\pm}, \mathsf{m}_{1\pm}]\subset \mathsf{m}_{2\pm}\,, \quad\quad [\mathsf{m}_{2\pm}, \mathsf{m}_{2\pm}]=0\,,\\ \label{commrels}
&[\mathsf{m}_{2-}, \mathsf{m}_{1+}]\subset \mathsf{m}_{1-}\,, \quad\quad [\mathsf{m}_{2+}, \mathsf{m}_{1-}]\subset \mathsf{m}_{1+}\\ \nonumber
&[\mathsf{m}_{1+}, \mathsf{m}_{1-}]\subset \mathsf{o}(n-2m)\oplus \mathsf{u}(m)\,,\quad\quad [\mathsf{m}_{2+}, \mathsf{m}_{2-}]\subset \mathsf{u}(m)\,.
\end{empheq}
In particular, the first line implies that the complex structure on $\mathsf{m}$, defined by the choice of holomorphic subspaces as in~(\ref{m2hol}) and~(\ref{m1hol}), is integrable.

Now, consider the group element $g\in \tO(n)$, and the standard Maurer-Cartan current
\bea
J=-g^{-1}dg=J_{\mathsf{h}}+J_{\mathsf{m}}\,.
\eea
According to~(\ref{mirrepdec}), we decompose the current as $J_{\mathsf{m}}=J_{\mathsf{m}_1}+J_{\mathsf{m}_2}$. The most general invariant metric then has the form
\bear
&&ds^2={a\over 2}\,\mathrm{Tr}(J_{\mathsf{m}_1}^2)+{b\over 2}\,\mathrm{Tr}(J_{\mathsf{m}_2}^2)=\\
&&= a\,\mathrm{Tr}(J_{\mathsf{m}_{1+}}J_{\mathsf{m}_{1-}})+b\,\mathrm{Tr}(J_{\mathsf{m}_{2+}}J_{\mathsf{m}_{2-}})\,.
\eear
In passing to the second line, we have taken into account that $\mathsf{m}_{1\pm}$ and $\mathsf{m}_{2\pm}$ are isotropic subspaces w.r.t. the trace form, as one easily sees from~(\ref{m2hol}) and~(\ref{m1hol}). We thus see that the metric is Hermitian w.r.t. the chosen integrable complex structure. The corresponding fundamental Hermitian form is, by definition,
\bea
\Omega={a\over 2}\,\mathrm{Tr}\left(J_{\mathsf{m}_{1+}}\wedge J_{\mathsf{m}_{1-}}\right)+{b\over 2}\,\mathrm{Tr}\left(J_{\mathsf{m}_{2+}}\wedge J_{\mathsf{m}_{2-}}\right)\,.
\eea
Let us check, in what case this form is closed, i.e. in what case the metric is K\"ahler. We  use the Maurer-Cartan equation $dJ-J\wedge J=0$ together with the commutation relations~(\ref{commrels}) of the algebra to obtain
\bear
d\Omega={a\over 2}\,\mathrm{Tr}\left(\{J_{\mathsf{m}_{2+}}, J_{\mathsf{m}_{1-}}\}\wedge J_{\mathsf{m}_{1-}}\right)-{a\over 2}\,\mathrm{Tr}\left(J_{\mathsf{m}_{1+}}\wedge \{J_{\mathsf{m}_{2-}}, J_{\mathsf{m}_{1+}}\}\right)+\\ \nonumber
+ {b\over 4}\,\mathrm{Tr}\left(\{J_{\mathsf{m}_{1+}}, J_{\mathsf{m}_{1+}}\}\wedge J_{\mathsf{m}_{2-}}\right)-{b\over 4}\,\mathrm{Tr}\left(J_{\mathsf{m}_{2+}}\wedge \{ J_{\mathsf{m}_{1-}}, J_{\mathsf{m}_{1-}}\}\right)\,.
\eear
One finds that $d\Omega=0$ in one of the three cases:
\begin{itemize}
\item $2a=b$\,,
\item $\mathsf{m}_{1}=0$, so that $ds^2={b\over 2}\,\mathrm{Tr}(J_{\mathsf{m}_2}^2)$\,,
\item $\mathsf{m}_{2}=0$, so that $ds^2={a\over 2}\,\mathrm{Tr}(J_{\mathsf{m}_1}^2)$\,.
\end{itemize}
As mentioned earlier, the last two cases correspond to symmetric spaces. Finally, we note that the metric featuring in our sigma models is the so-called normal (reductive) metric, defined by $ds^2={a\over 2}\,\mathrm{Tr}(J_{\mathsf{m}}^2)$. When viewed as part of the general family of metrics above, it corresponds to $b=a$. It follows that this metric is K\"ahler only in the case of symmetric spaces.

\subsection{Symplectic Grassmannians $\mathsf{SGr}(m, 2n)$.}\label{symplalgapp} 

In this Appendix we construct invariant metrics on symplectic Grassmannians. Most of the methods can be easily adapted from the orthogonal Grassmannian theory presented above. First we recall the definition of the symplectic algebra $\mathsf{sp}(2n)$. It is comprised of transformations preserving the symplectic form on $\CC^{2n}$. We can take $\mathcal{J}$ of~(\ref{compstruct}) as our model symplectic form:
\bea
\mathcal{J}=\mathds{1}_n\otimes i\sigma_2\,.
\eea
The symplectic algebra is then
\bea
\mathsf{sp}(2n)=\Bigl\{\quad \tau\in \mathrm{Mat}_{2n}(\CC):\quad\quad \tau^t \mathcal{J}+\mathcal{J} \tau=0\quad  \Bigl\}\,.
\eea
One can take the following basis:
\bear
&&\mathsf{sp}(2n)=\mathrm{Span}\left(\alpha_1 \otimes \mathds{1}_2,\quad \alpha_2 \otimes i\sigma_2,\quad \beta_1\otimes \sigma_1, \quad \beta_2\otimes \sigma_3\right)\,,\\ \nonumber 
&&\alpha_1^t=-\alpha_1,\quad \alpha_2^t=\alpha_2,\quad \beta_1^t=\beta_1,\quad \beta_2^t=\beta_2\,.
\eear
Notice that it is parallel to the choice of basis~(\ref{alphadef})-(\ref{betadef}), except for a different choice of symmetry properties for $\beta$.

\begin{figure}
\begin{center}
 \begin{tikzpicture}[every node/.style={text height=3.5ex,text width=2em}]
\put(-135,-7) {$\mathsf{sp(2n)}=$};
\put(-60,20) {$\mathsf{u}(m)\oplus \mathsf{m_2}$};
\put(5,-32) {\small $\mathsf{sp}(2n-2m)$};
\put(10,20) {$A\in\mathsf{m_1}$};
\put(-60,-32) {$-A^t\in\mathsf{m_1}$};
\matrix[matrix of math nodes,
        left delimiter=(,
        right delimiter=),
        nodes in empty cells] (m)
{
       &  &   &    \\
  &         &   &             \\
       &   &   &             \\
       &   &   &             \\
};
\path[draw, line width = 0.3mm] (m-1-2.north east) -- (m-4-2.south east);
\path[draw, line width = 0.3mm] (m-2-1.south west) -- (m-2-4.south east);
    \end{tikzpicture}
    \end{center}
\caption{Decomposition~(\ref{spndecomp}) of the Lie algebra $\mathsf{sp(2n)}$. Any element of $\mathsf{m_1}$ is determined by $A\in\mathrm{Hom}(\CC^{2n-2m}, \CC^{2m})$.}
\label{spndecompfig}
    \end{figure}

According to the definition of the symplectic Grassmannian~(\ref{SGr}), we perform the Lie algebra decomposition
\bear\label{spndecomp}
&&\mathsf{sp}(2n)=\mathsf{h}\oplus \mathsf{m}\,,\quad\quad \textrm{where}\quad\quad \\ \nonumber
&&\mathsf{h}=\mathsf{u}(m)\oplus \mathsf{sp}(2n-2m)\,,\quad\quad \mathsf{m}=\mathsf{h}^{\perp}\,,
\eear
where by $\mathsf{h}^{\perp}$ we mean the orthogonal complement to $\mathsf{h}$  w.r.t. the Killing metric. Subalgebra $\mathsf{sp}(2n-2m)\subset \mathsf{sp}(2n)$ is embedded diagonally. As for $\mathsf{u}(m)$, recall that
\bea
\mathsf{u}(m)=\mathsf{o}(2m)\; \bigcap \; \mathsf{sp}(2m)\,.
\eea
Thus, $\mathsf{u}(m)$ is embedded in $\mathsf{sp}(2m)$ by the same formula~(\ref{alphadef}):
\bea\label{alphadef2}
\alpha=\alpha_1\otimes \mathds{1}_2+\alpha_2\otimes i\sigma_2\,\quad\quad \textrm{where}\quad\quad \alpha_1^t=-\alpha_1,\quad \alpha_2^t=\alpha_2\,.
\eea
Again, as in the orthogonal case, over the real numbers $\mathsf{m}$ is split into two irreducible representations ($[\mathsf{h}, \mathsf{m}_i]\subset \mathsf{m}_i$):
\bea\label{mirrepdec2}
\mathsf{m}=\mathsf{m}_1\oplus \mathsf{m}_2\,.
\eea
This decomposition is shown in Fig.~\ref{spndecompfig}. 
In particular, $\mathsf{m}_2$ is furnished by the matrices 
\bea\label{betadef2}
\beta=\beta_1\otimes \sigma_1+\beta_2\otimes \sigma_3\,\quad\quad \textrm{where}\quad\quad \beta_1^t=\beta_1,\quad \beta_2^t=\beta_2\,.
\eea
One can define a complex structure on $\mathsf{m}_1$ and $\mathsf{m}_2$ in the same way as in the orthogonal case:
\bear\label{m2holsp}
&&(\mathsf{m}_2)_{\pm}=\Bigl\{\quad \beta\otimes (\sigma_1\mp i \sigma_3)\,,\quad\quad \beta^t=\beta\quad\Bigl\}\,, \\
\label{m1holsp}
&&(\mathsf{m}_1)_{\pm}=\Biggl\{\quad v\otimes \begin{pmatrix}1\\ \pm i\end{pmatrix}\quad\Biggl\}\,.
\eear
All subsequent calculations reproduce the ones of the previous Appendix. As a result, one arrives at the same commutation relations~(\ref{commrels}), up to the evident change~$\mathsf{o}\to \mathsf{sp}$:
\begin{empheq}[box = \fcolorbox{black}{white}]{align}
\nonumber
&[\mathsf{m}_{2\pm}, \mathsf{m}_{1\pm}]=0\,,\quad\quad [\mathsf{m}_{1\pm}, \mathsf{m}_{1\pm}]\subset \mathsf{m}_{2\pm}\,, \quad\quad [\mathsf{m}_{2\pm}, \mathsf{m}_{2\pm}]=0\,,\\ \label{commrels2}
&\,[\mathsf{m}_{2-}, \mathsf{m}_{1+}]\subset \mathsf{m}_{1-}\,, \quad\quad [\mathsf{m}_{2+}, \mathsf{m}_{1-}]\subset \mathsf{m}_{1+}\,,\\ \nonumber
&\, [\mathsf{m}_{1+}, \mathsf{m}_{1-}]\subset \mathsf{sp}(n-2m)\oplus \mathsf{u}(m)\,,\quad\quad [\mathsf{m}_{2+}, \mathsf{m}_{2-}]\subset \mathsf{u}(m)\,.
\end{empheq}
Again, the most general invariant metric  has the form
\bear
&ds^2=a\,\mathrm{Tr}(J_{\mathsf{m}_{1+}}J_{\mathsf{m}_{1-}})+b\,\mathrm{Tr}(J_{\mathsf{m}_{2+}}J_{\mathsf{m}_{2-}})\,,
\eear
where $J_{\mathsf{m}_\pm}$ are components of the Maurer-Cartan current. The metric is K\"ahler if $2a=b$, or if the space is symmetric (when $\mathsf{m}_{1}=0$). The metrics featuring in our sigma models are the reductive metrics $ds^2={a\over 2} \mathrm{Tr}(J_{\mathsf{m}}^2)$, and it follows that these are K\"ahler only in the  symmetric space case (which is the case of Lagrangian Grassmannian).

\section{The generalized $\alpha$-gauge}\label{alphagaugeapp}

In the main text we concluded that, in order to integrate over the $V$ and $\thickbar{V}$ fields, in models of orthogonal and symplectic Grassmanians one needs to use the $\alpha$-gauge trick due to the degeneracy of the corresponding quadratic form. For the simplification of calculations we picked a convenient value of $\alpha$ prior to inverting the quadratic form. A more rigorous approach would be to  prove that sigma model metrics do not depend on the gauge-fixing parameter. In the present section we invert the quadratic form for arbitrary value of $\alpha$ in the case of orthogonal Grassmanians, proving that the parameter drops out in the final expression for the metric. In the symplectic case calculations are similar. 

The interaction part of Lagrangian (\ref{SOgaugefixed}) in an arbitrary $\alpha$-gauge with normalization condition ${\thickbar{U}U=\mathds{1}_m}$ is given by\footnote{Here we add to the Lagrangian a gauge-fixing term $\mathrm{Tr}(\mathcal{F} \thickbar{\mathcal{F}})$ with coefficient $\vkappa/4\pi\alpha$, thus slightly changing normalization of $\alpha$ for the convenience of calculations.}
\bear
&&\mathcal{L}_{int}=\frac{\vkappa}{4\pi}\,\mathrm{Tr}(J_{\mathsf{o}} \thickbar{J}_{\mathsf{o}})+\frac{\vkappa}{4\pi\alpha}\mathrm{Tr}(\mathcal{F} \thickbar{\mathcal{F}})=\\ &&=2\pi\vkappa\,\mathrm{Tr}\left[\thickbar{V}V-
\left(1-\frac{1}{\alpha}\right)UV\thickbar{U}^t\thickbar{V}^t+\frac{1}{\alpha}\thickbar{V}V\thickbar{U}^tU^t\right].
\eear
Here the $U$ and $V$ variables are entangled; it is convenient to write this quadratic form in terms of matrix components:
\begin{equation}
\mathcal{L}_{int}=2\pi\vkappa\,V_{\alpha a}\left[\delta_{\alpha\beta}\delta_{ab}-\left(1-\frac{1}{\alpha}\right)\thickbar{U}_{\beta a}U_{b\alpha}+\frac{1}{\alpha}\delta_{\alpha\beta}\thickbar{U}_{\gamma a}U_{b\gamma}\right]\thickbar{V}_{b\beta}:=V_{\alpha a}\,K_{ab,\alpha\beta}\,\thickbar{V}_{b\beta},
\end{equation}
where coefficients of the quadratic form are denoted by $K_{ab,\alpha\beta}$. 
Making the most general ansatz $$K_{ab,\alpha\beta}^{-1}=\frac{1}{2\pi\vkappa}\left(\delta_{\alpha\beta}\delta_{ab}-A\,\thickbar{U}_{\beta a}U_{b\alpha}-B\,\delta_{\alpha\beta}\thickbar{U}_{\gamma a}U_{b\gamma}\right)$$ for the inverse quadratic form and solving linear equations for $A$ and $B$, one finds $$A=\frac{1-\alpha}{4},\quad B=\frac{3-\alpha}{4}\,.$$
Returning to matrix notation, one can write the sigma model Lagrangian upon elimination of the $V$ and $\thickbar{V}$ fields as follows:
\begin{align}
&\mathcal{L}_{SM}={4\over 2\pi\vkappa}\,\mathrm{Tr}\left[D \thickbar{U}\,\thickbar{D}U\right]
-{2A\over \pi\vkappa}\mathrm{Tr}\left[U\left(D\thickbar{U}\right)\mathds{h}_n\,\thickbar{U}^t\left(\thickbar{D}U\right)^t\mathds{h}_n\right]-\\ \nonumber
&-{2B\over \pi\vkappa}\mathrm{Tr}\left[\mathds{h}_n\left(D
\thickbar{U}\right)^t\left(\thickbar{D}\,U\right)^t\mathds{h}_n\left(U \thickbar{U}\right)\right]+\mathrm{Tr}\left(\mathcal{A}_+ U^t \mathds{h}_n U-\thickbar{\mathcal{A}_+}\thickbar{U} \mathds{h}_n\thickbar{U}^t\right).
\end{align}
Due to the constraints one can drop the gauge fields $\mathcal{A}$ in  the second and third terms. Besides, using the derivative of the constraint $\thickbar{U}\mathds{h}_n\thickbar{U}^t=0$,  one can show that dependence on $\alpha$ falls out, 
\begin{align}
A\,&\mathrm{Tr}\left[U\left(D\thickbar{U}\right)\mathds{h}_n\,\thickbar{U}^t\left(\thickbar{D}U\right)^t\mathds{h}_n\right]
+B\,\mathrm{Tr}\left[\mathds{h}_n\left(D
\thickbar{U}\right)^t\left(\thickbar{D}\,U\right)^t\mathds{h}_n\left(U \thickbar{U}\right)\right]=\notag\\
&={1\over 2}\mathrm{Tr}\left[\left(\partial
\thickbar{U}\right)^t\left(\,\thickbar{\partial}\,U\right)^t\mathds{h}_n\left(U \thickbar{U}\right)\mathds{h}_n\right].
\end{align}
It means that this Lagrangian completely coincides with (\ref{orthomnlagr}).

\section{$\beta$-function for $\CP^1$ target spaces}\label{CP1betaapp}

Here we check our $\beta$-function calculations by noting that each family~(\ref{UGr})-(\ref{SGr}) contains at least one special case when the target space is a sphere $\CP^1$. Indeed,
\bea
\mathsf{Gr}(1,2)\simeq \mathsf{OGr}(1, 3)\simeq \mathsf{OGr}^+(2, 4)\simeq \mathsf{SGr}(1, 2)\simeq \CP^1\,.
\eea
We list the relevant values of the $\beta$-functions for these cases:
\begin{align}
&-\frac{\beta^{\mathsf{Gr}}_\vkappa\big|_{n=2}}{\vkappa^2}=2\,,\qquad -\frac{\beta^{\mathsf{OGr}}_\vkappa\big|_{n=3}}{\vkappa^2}=1\,, \qquad-\frac{\beta^{\mathsf{OGr}}_\vkappa\big|_{n=4}}{\vkappa^2}=2\,,\qquad -\frac{\beta^{\mathsf{SGr}}_\vkappa\big|_{n=1}}{\vkappa^2}=4\,.\label{betacp11}
\end{align}

\begin{itemize}
\item We start with $\mathsf{Gr}(1,2)$, which will serve as our reference case. Here $U$ is a vector in $\CC^2$, and the Lagrangian is
\bea
\mathcal{L}=2\left(V\thickbar{D}U+\thickbar{U}\,D\thickbar{V}\right)+2\pi\vkappa \,\left(\thickbar{U}U\right)\cdot \left(V\thickbar{V}\right)\,.
\eea
Variation w.r.t. the gauge field $\thickbar{\mathcal{A}}$ gives the constraint $U_1 V_1+U_2 V_2=0$. Using $\CC^\ast$ gauge symmetry, we may set $U_1=1$ and  solve the constraint as $V_1=-U_2 V_2$. Integrating over the remaining momentum $V_2$, we get
\bea\label{CP1Lagr}
\mathcal{L}={4\over 2\pi\vkappa}\,\frac{\big|\,\thickbar{\dd}U_2\big|^2}{(1+|U_2|^2)^2}\,.
\eea
We will thus aim to get the same Lagrangian in all remaining cases.

\item The next case is $\mathsf{OGr}(1, 3)$. Here $U$ is a vector in $\CC^3$ satisfying the constraint 
\bea
\sum_{i=1}^3 U_i U_i=0\,.
\eea
The relevant Lagrangian is
(\ref{orthomnlagr}) where the second term vanishes due to the constraint.
Using homogeneous coordinates, one has
\bea\label{LagrOGr13}
\mathcal{L}={4\over 2\pi\vkappa\,|U|^2}\,\left(\big|\,\thickbar{\dd}U\big|^2-\frac{|\thickbar{U}\,\thickbar{\dd}U|^2}{|U|^2}\right)\,.
\eea
We may now resolve the constraint via
\bea\label{Veronese}
U_1=u_1^2+u_2^2\,,\quad\quad U_2=i(u_1^2-u_2^2)\,,\quad\quad U_3=2i \,u_1 u_2\,.
\eea
This is the so-called Veronese embedding $\CP^1\hookrightarrow \CP^2$. Using projective invariance, we may now set $u_1=1$. Substituting~(\ref{Veronese}) into the Lagrangian~(\ref{LagrOGr13}), we arrive at~(\ref{CP1Lagr}) with an extra factor of 2, 
\bea
\mathcal{L}={4\over \pi\vkappa}\,\frac{|\,\thickbar{\dd}u_2|^2}{(1+|u_2|^2)^2}\,.
\eea

\item $\mathsf{OGr}^+(2, 4)$. Here, according to~(\ref{orthomnlagr}) and~(\ref{maxOGrassform}), one should take a $2\times 4$-matrix $U=(u_1, u_2)$, and the Lagrangian is determined by the (generalized) Fubini-Study metric on the Grassmannian $\mathsf{Gr}(2,4)$. In homogeneous coordinates it reads
\bear\label{LagrOGr24}
\mathcal{L}={1\over \pi\vkappa}\,\mathrm{Tr}\left(\thickbar{D}U {1\over \thickbar{U}U} D\thickbar{U}\right)\,,\\
\textrm{where}\quad\quad \thickbar{D}U=\thickbar{\dd} U- U\,\thickbar{\mathcal{A}}\,.
\eear
One also has the isotropy constraints $U^tU=0$, or, more explicitly,  $u_1^2=u_2^2=u_1 u_2=0$. In other words, one has an embedding $\mathsf{OGr}(2, 4)\hookrightarrow \mathsf{Gr}(2,4)$.

The Lagrangian is explicitly gauge-invariant w.r.t. $\tGL(2, \CC)$ acting as $U\to Ug$. Using this gauge symmetry, as well as the conditions $u_1^2=u_2^2=0$, we may bring $U$ to the form
\bea\label{UsolG24}
U=\begin{pmatrix}
    2i\upalpha& 0\\
    0& 2i \upbeta\\
    i(\upalpha^2-1) & i (\upbeta^2-1)\\
    \upalpha^2+1 & \upbeta^2+1
\end{pmatrix}\,.
\eea
Here each column is parametrized as in~(\ref{Veronese}), where we have additionally passed to inhomogeneous coordinates. The remaining constraint $u_1u_2=0$ gives $\upalpha^2+\upbeta^2=0$, which has two solutions $\upbeta=\pm i \upalpha$. This reflects the fact that $\mathsf{OGr}(2, 4)$ has two connected components:
\bea
\mathsf{OGr}(2, 4)\simeq \CP^1 \bigsqcup \CP^1\,.
\eea
One can choose either of the two $\CP^1$'s, i.e. either solution $\upbeta=\pm i \upalpha$, and substitute~(\ref{UsolG24}) into the Lagrangian, at the same time eliminating the gauge field through its e.o.m. As a result, one gets (relabelling $\upalpha \to u$)
\bea
\mathcal{L}={4\over 2\pi\vkappa}\,\frac{|\,\thickbar{\dd}u|^2}{(1+|u|^2)^2}\,.
\eea
This completely coincides with (\ref{CP1Lagr}).

\item We finish with the elementary symplectic Grassmannian $\mathsf{SGr}(1, 2)$. Here $U$ lives in $\CC^2$ and the relevant Lagrangian, according to~(\ref{lagrsympl2})-(\ref{SGrassmaxform}), is
\bea\label{LagrSGr12}
\mathcal{L}={1\over \pi\vkappa\,|U|^2}\,\left(|\,\thickbar{\dd}U|^2-\frac{|\thickbar{U}\,\thickbar{\dd}U|^2}{|U|^2}\right)={1\over \pi\vkappa}\,\frac{|\,\thickbar{\dd}u|^2}{(1+|u|^2)^2}\,,
\eea
where in the last equality we passed over to the inhomogeneous coordinate. 
This differs from~(\ref{CP1Lagr}) by a factor of $1\over 2$.
\end{itemize}

The four special cases are thus in agreement with~(\ref{betacp11}).

\setstretch{0.8}
\setlength\bibitemsep{5pt}
\printbibliography

\end{document}